\documentclass[%
reprint,
superscriptaddress,
 amsmath,amssymb,
aps,
prb,
]{revtex4-2}

\usepackage{pifont}
\usepackage{verbatim}
\usepackage{subfigure}
\usepackage{graphicx}
\usepackage{dcolumn}
\usepackage{bm}
\usepackage{xcolor} 
\usepackage{hyperref}

\begin{document}

\title{Higher-order topological insulators in antiperovskites}

\author{Yuan Fang}
\affiliation{Department of Physics and Astronomy, Stony Brook University, Stony Brook, New York 11974, USA}

\author{Jennifer Cano}
\affiliation{Department of Physics and Astronomy, Stony Brook University, Stony Brook, New York 11974, USA}
\affiliation{Center for Computational Quantum Physics, The Flatiron Institute, New York, New York 10010, USA}

\date{\today}

\begin{abstract}
We predict that a family of antiperovskite materials realize a higher order topological insulator phase, characterized by a previously introduced $\mathbb{Z}_4$ index. A tight binding model and a $k\cdot p$ model are used to capture the physics of the bulk, surface and hinge states of these materials. A phase diagram of the higher order and weak topological invariants is obtained for the tight binding model. The mirror Chern number is also discussed. In order to reveal the gapless hinge states in the presence of mirror Chern surface states, several ways of opening the surface gap are proposed and confirmed by calculation, including cleaving the crystal to reveal a low-symmetry surface, building a heterostructure, and applying strain. Upon opening the surface gap, we are able to study the hinge states by computing the momentum space band structure and real space distribution of mid-gap states.
\end{abstract}

\maketitle

\section{Introduction\label{intro}}

The recent classification \cite{bradlyn2017topological,Po2017,Cano2018building,Vergniory2017graph,Bradlyn2018band,vergniory2019complete,Elcoro2017,zhang2019catalogue} of topological insulators with crystal symmetry has led to the discovery of a new type of topological phase, the higher order topological insulator (HOTI) \cite{Benalcazar2017science,Benalcazar2017electric,Khalaf2018symmetry,Khalaf2018higher,schindler2018higher,schindler2018higherBi,Song2017,Langbehn2017,Geier2018,Imhof2017,Peterson2018,SerraGarcia2018,Noh2018,Trifunovic2019higher}.
HOTIs in three dimensions (3D) are gapped in the bulk and on all surfaces, but have one dimensional gapless modes along ``hinges'' where two surfaces meet.
Here, we are concerned with HOTIs in 3D protected by time reversal and inversion symmetry: for these HOTIs, the one dimensional gapless hinge mode is a helical mode. Hence, when combined with superconductivity, HOTIs present a new route to engineering Majorana fermions from topological insulator heterostructures \cite{queiroz2019splitting,fu2008superconducting,nilsson2008splitting,fu2009josephson,tanaka2009manipulation,linder2010unconventional,PhysRevLett.121.196801}.

Realizing such a heterostructure requires a 3D HOTI material.
So far, Bismuth, which has a continuous direct band gap, is topologically equivalent to a HOTI \cite{schindler2018higherBi} and strained SnTe has also been predicted \cite{schindler2018higher}.
In addition, several weak TIs are predicted to be nontrivial HOTIs when their surfaces are gapped by breaking translation symmetry \cite{schindler2018higher,queiroz2019partial}.

\begin{figure}[t]
    \centering
    \includegraphics[width = 8.5cm]{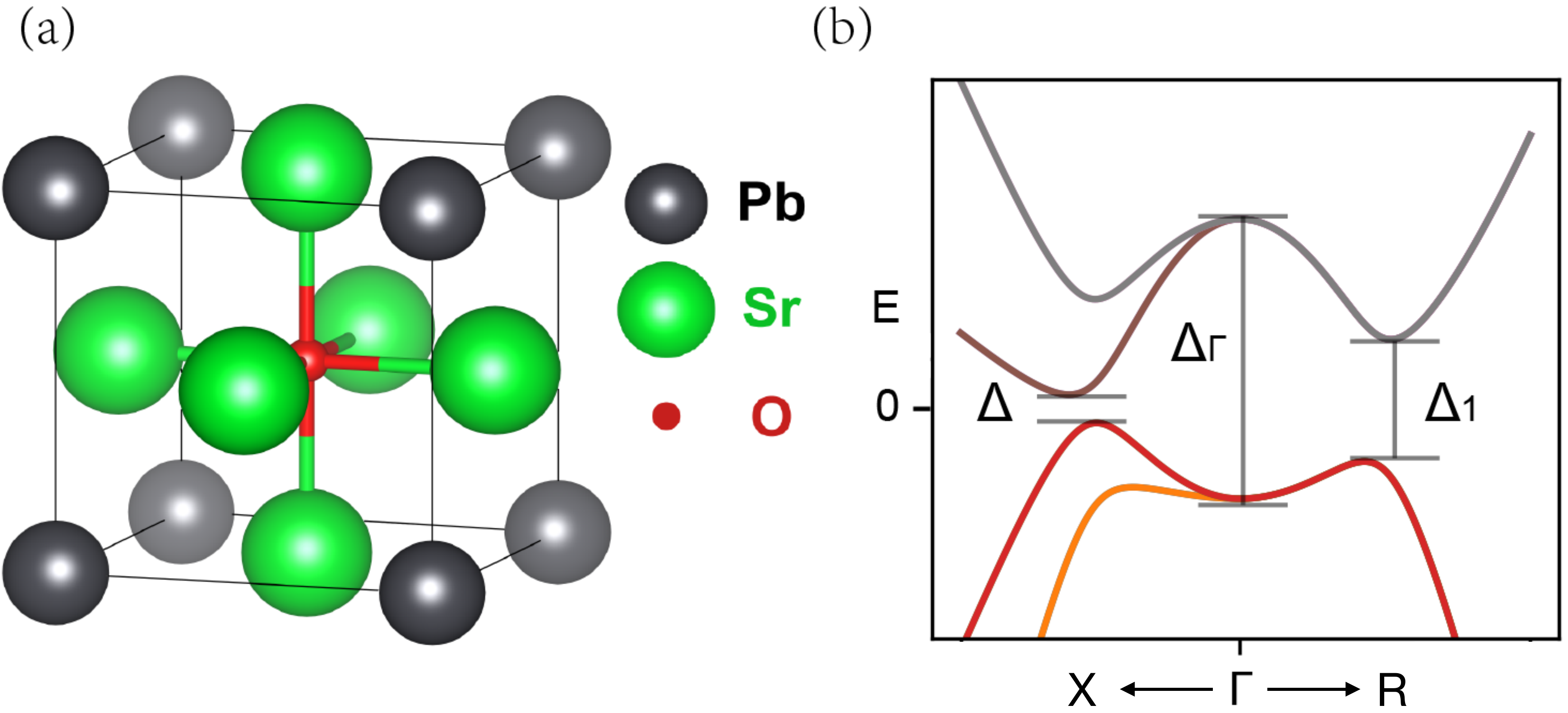}
    \caption{(a) Crystal structure of the antiperovskite Sr$_3$PbO, in space group $Pm\bar{3}m$ \cite{materialsproject}. The Sr, Pb, and O atoms are shown in green, black, and red, respectively. {(b) Schematic band structure of antiperovskites along the $(k,0,0)$ and $(k,k,k)$ axes. The smallest gap, $\Delta$, sits at an avoided crossing point along $\Gamma-X$ gapped by spin-orbit coupling}.} 
    \label{figcrystal}
\end{figure}

In this manuscript, we propose a family of HOTIs in the antiperovskites as a promising material class.
The antiperovskites are familiar to the topological community as mirror Chern insulators \cite{hsieh2014topological}. 
Many antiperovskites exhibit a ``double band inversion,'' caused by the inversion of two $J=3/2$ quartets \cite{Sun2010,kariyado2011three,Kariyado2012}, which results in a trivial $\mathbb{Z}_2$ index, but a nontrivial mirror Chern number. 
Here, we show that the double band inversion is exactly the necessary ingredient to realize the HOTI protected by inversion and time reversal symmetry. 
Using the Topological Materials Database \cite{vergniory2019complete}, we report eight compounds -- Ca$_3$SnO, Ca$_3$PbO, Ca$_3$GeO, Ba$_3$PbO, Sr$_3$PbO, Sr$_3$SnO, Sr$_3$BiN, and Ti$_3$TlN -- exhibiting a nontrivial HOTI index, with the largest bulk gaps in Sr$_3$PbO and Sr$_3$SnO greater than $50$ meV. 
Due to the double band inversion, the bulk gap, $\Delta$, sits at an avoided crossing along $\Gamma-X$, as shown schematically in Fig.~\ref{figcrystal}(b).

{We construct a minimal tight-binding model that goes beyond the continuum model in Ref.~\onlinecite{hsieh2014topological} in an essential way: it breaks the continuous rotational symmetry down to the crystal symmetry group so that surfaces with a trivial mirror Chern number are gapped (Sec.~\ref{Ham}).}
We compute a phase diagram for this model and show that it exhibits several different topological phases, including the HOTI phase realized by the antiperovskites (Sec.~\ref{index}).
We then consider the helical hinges modes in the HOTI phase.
For a finite-size sample with mirror-preserving surfaces, the hinge modes will be masked by the gapless mirror Chern surface states.
Thus, we propose to measure the hinge modes in samples where mirror symmetry is broken by either cleaving along a low-symmetry surface or by applying mirror-breaking strain (Sec.~\ref{mc}).
When the mirror surface states are gapped, we numerically confirm the presence of the hinge states (Sec.~\ref{hinge}).
Finally, in Sec.~\ref{mat}, we connect our model Hamiltonian to specific antiperovskite materials and show that they reside in the nontrivial topological phase.
We give an outlook in Sec.~\ref{outlook}.


\section{Tight binding Hamiltonian\label{Ham}}

Antiperovskites have the chemical formula A$_3$BX. Ideally, they form the primitive cubic structure shown in Fig.~\ref{figcrystal}, where the A atom is at the faces of the cube, the B atom is at the corners, and the X atom is at the center.
It was noted in previous work \cite{Sun2010,Kariyado2012,hsieh2014topological} that several of these compounds display a band structure with two band inversions, due to the spin-3/2 quartet of $d$ orbitals from the A atom inverting beneath the spin-3/2 quartet of $p$ orbitals from the B atom, and resulting in a topological crystalline insulator (previously noted for its mirror Chern number \cite{hsieh2014topological} and here noted for its nontrivial HOTI invariant.)

Hsieh, et al, introduced a linear order $k\cdot p$ model to capture the non-zero mirror Chern number of certain antiperovskites \cite{hsieh2014topological}. 
However, higher orders in $k$ are necessary to correctly model the surface states: in particular, the linear model has full ${SO}(3)$ rotation symmetry; consequently, surface states appear along every direction, including directions that do not correspond to a mirror plane of the crystal.
In addition, a $k.p$ model is inadequate to compute topological indices that rely on global properties of the band structure.
For these two reasons, we are motivated to build a tight-binding Hamiltonian.

The remainder of this section is devoted constructing the tight-binding model.
We consider antiperovskites in the space group $Pm\bar{3}m$; the crystal structure is shown in Fig.~\ref{figcrystal}. 
The primitive cubic space group is generated by the rotations $C_{2,001}$, $C_{2,010}$, $C_{2,110}$, and $C_{3,111}$ and inversion, $\cal P$. (We use $C_{n,ijk}$ to denote an $n$-fold rotation about the axis $i\hat{x} + j\hat{y} + k\hat{z}$.) We also enforce time reversal symmetry, $\cal T$.

The low-energy physics is derived from four $d$ and four $p$ orbitals, which form spin-$3/2$ quartets with opposite parity
\cite{Sun2010,kariyado2011three,Kariyado2012,hsieh2014topological}.
In the trivial case without band inversion, the $d$ orbitals form the conduction band and the $p$ orbitals form the valence band, while in the nontrivial case, the quartets invert at the origin.
We will build the most general short-range tight-binding model consistent with the symmetry and orbital content we have just described; 
therefore, many different topological and trivial phases are realized within this model, as summarized by the phase diagram in Fig.~\ref{pdfigure}.

We now describe how to construct the symmetry operators.
We use three copies of the Pauli matrices $\sigma_i$, $\tau_i$, $\rho_i$, $i=0,x,y,z$ that act on different degrees of freedom to describe the eight-dimensional Hamiltonian.
Matrix forms of the symmetry operations are obtained in the following way: the Pauli matrix $\rho$ labels the $d$ and $p$ quartets; therefore, the inversion (parity) operator is given by ${\cal P} = \sigma_{{0}} \tau_{{0}} \rho_{z}$.
The Pauli matrices $\sigma$ and $\tau$ together describe spins within the $J=3/2$ quartet, such that $J_z = \frac12\sigma_z\tau_0+\sigma_0\tau_z$; the matrix form of $\vec{J}=(J_x,J_y,J_z)$ is in Appendix~\ref{jmatrix}.
The operator for a rotation by angle $\theta$ about an axis $\hat{n}$ is given by $\exp{(-i \vec{J} \cdot \hat{n} ~\theta)}\rho_0$.
The matrices of the crystal symmetries and the time reversal operator are explicitly shown here: 
\begin{align}
    {\cal P} &= \sigma_0\tau_0\rho_z \label{eqn_inversion}\\
    {\cal T} &= -i\sigma_y\tau_x\rho_0 K \\
    C_{2,001} &= i\sigma_z\tau_0\rho_0, \\
    C_{2,010} &= -i\sigma_y\tau_x\rho_0, \\
    C_{2,110} &= \frac{i}{\sqrt{2}}(\sigma_x\tau_y+\sigma_y\tau_y)\rho_0, \\
    C_{3,111} &= -\frac14 ((\sigma_0+i\sigma_z)\tau_0+i(\sigma_x-\sigma_y)\tau_x \nonumber \\
     &+\sqrt{3}(\sigma_0-i\sigma_z)\tau_y+\sqrt{3}(\sigma_x+\sigma_y)\tau_z )\rho_0,\label{eqn_C3}
\end{align}
where $K$ is the complex conjugate operator. For simplicity, all $\sigma_0$, $\tau_0$ and $\rho_0$ symbols will be omitted in the following text. \par

We derive the most general quadratic $k.p$ model that satisfies the symmetries (\ref{eqn_inversion})--(\ref{eqn_C3}).
The quadratic terms go beyond Ref.~\onlinecite{hsieh2014topological} and play an important role in understanding the physics of the surface states, as we discuss in detail in Appendix \ref{surface_theory}. 
Extending the $k\cdot p$ model to the whole Brillouin zone yields a tight binding Hamiltonian.
The extension is not unique: we have chosen the simplest model that satisfies all the symmetries by including only nearest and next nearest hopping in position space. Written in momentum space, the Hamiltonian is given by four blocks:
\begin{widetext}
\begin{gather}
\label{Hameqn}
{\cal H}_{{k}}
 =
  \begin{bmatrix}
   {\cal H}_{{0k}}({m},\alpha_{1},\beta_{1},\gamma_{1}) &
   {\cal H}_{{1k}}({v}_{1},{v}_{2}) \\
   {\cal H}_{{1k}}({v}_{1},{v}_{2})&
   {\cal H}_{{0k}}({-m},\alpha_{2},\beta_{2},\gamma_{2})
   \end{bmatrix},
\end{gather}
where the off-diagonal blocks are given by:
\begin{equation}
\label{eqnH1}
    {\cal H}_{{1k}} =
    {v}_{1} \left(\sin{{k}_{x}} {J}_{x}+\sin{{k}_{y}} {J}_{y}+\sin{{k}_{z}} {J}_{z}\right)+
    {v}_{2} \left(\sin{{k}_{x}} \tilde{{J}}_{x}+\sin{{k}_{y}} \tilde{{J}}_{y}+\sin{{k}_{z}} \tilde{{J}}_{z}\right),
\end{equation}
and the diagonal blocks by:
\begin{align}
\label{eqnH0}
     {\cal H}_{{0k}} = &m+\frac{\alpha}{2}(3-\cos{k_x}-\cos{k_y}-\cos{k_z}) \nonumber \\
    &+\frac{2\beta}{3}(\cos{k_x}J_x\cdot\tilde{J}_x+\cos{k_y}J_y\cdot\tilde{J}_y+\cos{k_z}J_z\cdot\tilde{J}_z) \nonumber \\ 
    &+\frac{2\gamma}{\sqrt{3}}(\sin{k_x}\sin{k_y}\{J_x,\tilde{J}_y\}+\sin{k_y}\sin{k_z}\{J_y,\tilde{J}_z\}+\sin{k_z}\sin{k_x}\{J_z,\tilde{J}_x\}).
\end{align}
The spin-3/2 matrices $J$ and $\tilde{J}$ are given in Appendix \ref{jmatrix}. The relation $\{J_i,\tilde{J}_j\}=\{J_j,\tilde{J}_i\}, ~(i,j=x,y,z)$ ensures that Eq. (\ref{eqnH0}) is symmetric in permuting $k_x$, $k_y$ and $k_z$.
For convenience, we provide the explicit expression of ${\cal H}_{0k}$:
\begin{align}
\label{eqnH0_1}
    {\cal H}_{{0k}} = &m+\frac{\alpha}{2}(3-\cos{k_x}-\cos{k_y}-\cos{k_z}) \nonumber \\
    &+\frac{\beta}{4} (\cos{k_x}+\cos{k_y}-2\cos{k_z})\sigma_z\tau_z
    +\frac{\sqrt{3}}{4}\beta(\cos{k_y}-\cos{k_x})\tau_x   \nonumber \\
    &+\gamma \sin{k_x}\sin{k_y}\sigma_x\tau_z+
    \gamma \sin{k_y}\sin{k_z}\sigma_y\tau_z+
    \gamma \sin{k_x}\sin{k_y}\tau_y.
\end{align}

\end{widetext}

The two $4\times 4$ blocks of the Hamiltonian (\ref{Hameqn}) correspond to the quartets of $d$ and $p$ orbitals, each separately described by ${\cal H}_{{0k}}$, but with opposite sign of the mass, $m$.
For certain values of $m$, the two groups of four bands invert at $\Gamma$.
This double band inversion does not change the $\mathbb{Z}_2$ topological index, but can drive the system into a topological crystalline phase, as shown in Fig.~\ref{pdfigure} and analyzed in Sec.~\ref{index}.
(This topological crystalline phase is responsible for both the mirror Chern number introduced in \cite{hsieh2014topological} as well as the HOTI phase discussed in this work.)

The off-diagonal blocks containing ${\cal H}_{{1k}}$ couple the $p$ and $d$ orbitals and open the bulk gap following a band inversion. A gap-closing topological phase transition occurs when the band crossing between the inverted $d$ and $p$ orbitals occurs at a high-symmetry point: in this case, ${\cal H}_{{1k}}$ cannot open the gap because ${\cal H}_{{1k}}$ vanishes at $k_i = 0,~ \pi$.
We will discuss these topological phase transitions in Sec.~\ref{index}. 
When the band structure is gapped, we will compute topological invariants of the four occupied bands.

Our model is sufficiently general to capture many different phases consistent with the orbitals and symmetry that we have described. We will apply it to specific materials in Sec.~\ref{mat}.

\section{$\mathbb{Z}_2$ and $\mathbb{Z}_4$ topological indices\label{index}}

We will now review the $\mathbb{Z}_2$ and $\mathbb{Z}_4$ topological indices and compute a phase diagram of these indices for the Hamiltonian (\ref{Hameqn}).

\subsection{Definition of $\mathbb{Z}_2$ and $\mathbb{Z}_4$ topological indices \label{z2z4}}
\begin{figure}
    \centering
    \includegraphics[width = 7cm]{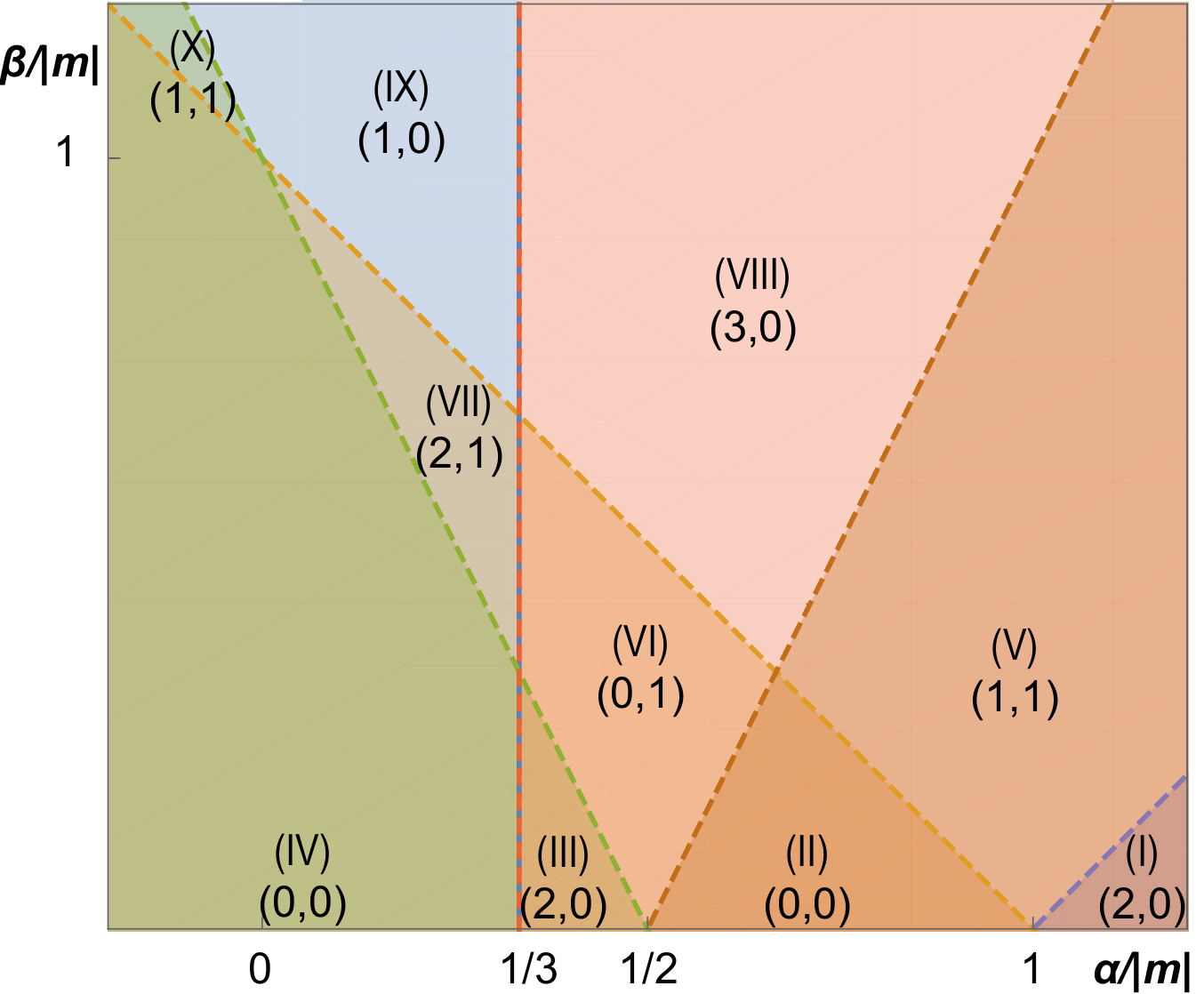}
    \caption{Phase diagram of the $\mathbb{Z}_4$ index and weak topological invariants for insulators. Ten regions labeled by Roman numerals are assigned a pair $(\kappa,\nu_w)$ indicating the $\mathbb{Z}_4$ index and weak topological invariant, respectively. (Recall that the $C_3$ symmetry makes all three weak topological invariants equal, so that we can indicate all three by one number, $\nu_w$.) $\kappa = 0$, $\nu_w = 0$ corresponds to a trivial insulator; $\kappa=1,3$ corresponds to a strong topological insulator; $\kappa=2$ corresponds to a higher order topological insulator (which may or may not have a nontrivial weak TI index); and $\nu_w = 1$ corresponds to a weak topological insulator. As discussed in Sec.~\ref{index}, an insulating phase can only result when $\alpha_1$ and $\beta_1$ are in the same region as $-\alpha_2$ and $-\beta_2$; hence, the axis labels $\alpha$ and $\beta$ are simultaneously indicating $\alpha_{1,2}$ and $\beta_{1,2}$.}
    \label{pdfigure}
\end{figure}

Recently, it was shown in Ref.~\cite{Khalaf2018symmetry} that in addition to the weak and strong topological invariants \cite{Fu2007topological,Moore2007topological,Roy2009topological}, the eigenvalues of the parity operator at time-reversal-invariant-momenta (TRIM) can be used to compute a $\mathbb{Z}_4$ topological index,
\begin{equation}
    \kappa = \frac{1}{4} \sum_{K \in \operatorname{TRIMs}}\left(n_{K}^{+}-n_{K}^{-}\right) \mod 4
    \label{Z4}
\end{equation}
where $n_K^{+}$/$n_K^{-}$  is the number of occupied states with even/odd parity. The familiar strong topological invariant $\nu_0$ is related to $\kappa$ by $\nu_0 = \kappa \mod 2$ \cite{Khalaf2018symmetry}. 
Thus, $\kappa = 1,~3$ indicates a strong TI. 
Further, $\kappa = 0$ indicates a trivial insulator. 
The HOTI phase, which is the focus of this paper, is realized when $\kappa=2$. 
(Note that there are many different types of HOTI phases, depending on the dimension and crystal symmetry, but here we will use HOTI to specifically refer to the three-dimensional phase protected by inversion and time reversal symmetry.)

The $\kappa = 2$ phase is also consistent with a weak topological insulator protected by translation and time reversal, which can be diagnosed by its inversion eigenvalues \cite{Fu_2007}. 
The weak and strong topological phases are captured by four $\mathbb{Z}_2$ invariants, usually denoted $(\nu_0;\nu_1,\nu_2,\nu_3)$ \cite{Fu2007topological,Moore2007topological,Roy2009topological}. 
Due to the cubic symmetry of our model, the three weak topological invariants are always equal; we denote them by $\nu_w \equiv \nu_1 = \nu_2 = \nu_3$, which can be computed by:
\begin{equation}
    \nu_w = \frac14 \sideset{}{'}\sum_{K \in \operatorname{TRIMs}}\left(n_{K}^{+}-n_{K}^{-}\right) \mod 2
    \label{Z2}
\end{equation}
where the prime on the summation symbol indicates that the sum should be restricted to TRIM within one of the time-reversal-invariant planes given by $k_i=\pi$; 
$n_K^{+}$/$n_K^{-}$ again indicates the number of occupied states with even/odd parity.\par

\subsection{Phase diagram of $\mathbb{Z}_2$ and $\mathbb{Z}_4$ indices}
Now let us return to our model. 
Fig.~\ref{pdfigure} shows the phase diagram of $\kappa$ and $\nu_w$ for all insulating phases of the Hamiltonian (\ref{Hameqn}). 
While the Hamiltonian contains many parameters, the $\gamma$ term and $v_{1,2}$ terms vanish at the TRIM points, where all $k$-components are equal to 0 or $\pi$. Therefore, only five parameters, $m$, $\alpha_{1,2}$ and $\beta_{1,2}$, can contribute to the phase diagram. 
Further, by enforcing that the system be an insulator and fixing $m$ as an overall energy scale, we can determine the topological indices $\kappa$ and $\nu_w$ of all insulating phases from only two parameters, as we now explain.\par

\begin{table}[b]
\caption{\label{table1} Occupancy of states with $+1$ parity eigenvalue at TRIM points. ``$+$'' means occupied (energy is negative relative to Fermi energy); ``$-$'' means unoccupied. Each sign ``$+$'', ``$-$'' has a two-fold Kramers degeneracy, so that the two signs represent four bands. To form an insulator, there must be four bands filled at each k-point. So every unoccupied state in this table corresponds to an occupied state from the bands with parity $-1$. In this sense, this table is also showing the parity of occupied states at TRIM points by regarding ``$-$'' and ``$+$'' to be parity eigenvalues. The symmetry-inequivalent TRIM points are $\Gamma=(0,0,0)$, $X=(\pi,0,0)$, $M=(\pi,\pi,0)$ and $R=(\pi,\pi,\pi)$.}
\begin{ruledtabular}
\begin{tabular}{ccccc}
{Phase}&
{$\Gamma$}&
{X}&
{M}&
{R}\\
\colrule
(I)&$+$ $+$ & $-$ $-$ & $-$ $-$ & $-$ $-$\\
(II)&$+$ $+$ & $+$ $+$ & $-$ $-$ & $-$ $-$\\
(III)&$+$ $+$ & $+$ $+$ & $+$ $+$ & $-$ $-$\\
(IV)&$+$ $+$ & $+$ $+$ & $+$ $+$ & $+$ $+$\\
(V)&$+$ $+$ & $-$ $+$ & $-$ $-$ & $-$ $-$\\
(VI)&$+$ $+$ & $-$ $+$ & $-$ $+$ & $-$ $-$\\
(VII)&$+$ $+$ & $-$ $+$ & $-$ $+$ & $+$ $+$\\
(VIII)&$+$ $+$ & $+$ $+$ & $-$ $+$ & $-$ $-$\\
(IX)&$+$ $+$ & $+$ $+$ & $-$ $+$ & $+$ $+$\\
(X)&$+$ $+$ & $-$ $+$ & $+$ $+$ & $+$ $+$
\end{tabular}
\end{ruledtabular}
\end{table}

First, without loss of generality, we can rescale all parameters by $|m|$, which sets the scale of the band gap at $\Gamma$, and take $m<0$.
Since ${\cal H}_{0k}(-m,\alpha,\beta,\gamma) = -{\cal H}_{0k}(m,-\alpha,-\beta,-\gamma)$, the phase diagram for $m>0$ can be obtained from that of $m<0$ by reversing the sign of $\alpha, \beta$ (recall the $\gamma$ term disappears at TRIM points) and swapping the occupied and unoccupied bands.

We now consider only the upper $4\times 4$ block of our eight-band model; the Hamiltonian of this block is given by ${\cal H}_{0k}(m,\alpha_1,\beta_1,\gamma_1)$.
Since this block describes $d$ orbitals, the inversion eigenvalues of these bands will always be $+1$, as can be checked from the inversion operator in (\ref{eqn_inversion}).
Choosing the Fermi energy to be $E_f = 0$, we now ask how many of these bands are occupied.
There are ten different cases, corresponding to the ten regions in Fig.~\ref{pdfigure}. 
The number of occupied $d$ bands at each TRIM point in each region is listed in Table \ref{table1}, where ``$+$'' indicates an occupied Kramers pair of $d$ orbitals and ``$-$'' indicates an unoccupied Kramers pair of $d$ orbitals.
Since there are four $d$ orbitals total, each TRIM point can be labeled by two $\pm$ signs in each region.

Now consider the lower $4\times 4$ block of (\ref{Hameqn}), given by ${\cal H}_{0k}(-m,\alpha_2,\beta_2,\gamma_2) = -{\cal H}_{0k}(m,-\alpha_2,-\beta_2,-\gamma_2)$.
The occupied/unoccupied bands are again described by Table~\ref{table1} by swapping the signs in the table (i.e., swapping the occupied/unoccupied bands), as well as swapping the signs of $\alpha_2,\beta_2$ (recall that $\gamma_2$ does not enter the calculation of the topological indices).

Since $\mathbb{Z}_2$ and $\mathbb{Z}_4$ are only defined for insulators, we now assume that our system is in an insulating phase.
If the system is insulating, there must be four occupied and four unoccupied bands at each TRIM point (since the bands at $\Gamma$ are fourfold degenerate due to the point group symmetry, it is not possible to have an insulator at any other filling.)
Since no two entries of Table~\ref{table1} are the same, it is only possible for the system to be insulating when $(\alpha_1,\beta_1)$ are in the same parameter regime as $(-\alpha_2,-\beta_2)$.
Conversely, if $(\alpha_1,\beta_1)$ and $(-\alpha_2,-\beta_2)$ are in different regimes, the system must be metallic. For example, if $(\alpha_1,\beta_1)$ are in phase (I) and $(-\alpha_2,-\beta_2)$ are in phase (II), then there will be four occupied bands at $\Gamma$ but no occupied band at $X$, so the system must be metallic.
Thus, the phase diagram of insulating phases can be deduced from only two parameters, $\alpha/|m|$ and $\beta/|m|$, which simultaneously indicate the parameter regime (shown in Fig.~\ref{pdfigure}) of $(\alpha_1,\beta_1)$ and $(-\alpha_2, -\beta_2)$.

Once the region is identified, the topological indices can be straightforwardly computed using Eqs.~(\ref{Z4}) and (\ref{Z2}) because Table~\ref{table1} not only indicates the occupied/unoccupied $d$ orbitals, but also the inversion eigenvalues of the occupied Kramers pairs. For example, two occupied Kramers pairs of $d$ orbitals is indicated by $++$ in Table~\ref{table1}; at the same time, if our eight-band model is insulating, then the $p$ orbitals must not be occupied and hence the inversion eigenvalues of the occupied Kramers pairs must also be $++$.
The topological indices for all regions are shown in Fig.~\ref{pdfigure} and also listed in Table \ref{table2}.
\par

\begin{table}[b]
\caption{\label{table2} Table of ten phases and the $\mathbb{Z}_4$ index $\kappa$ and weak topological invariant $\nu_w$.}
\begin{ruledtabular}
\begin{tabular}{ccccccccccc}
{phase}&
{(I)}&
{(II)}&
{(III)}&
{(IV)}&
{(V)}&
{(VI)}&
{(VII)}&
{(VIII)}&
{(IX)}&
{(X)}\\
\colrule
{$\kappa$}&
2&
0&
2&
0&
1&
0&
2&
3&
1&
1\\
{$\nu_w$}&
0&
0&
0&
0&
1&
1&
1&
0&
0&
1\\
\end{tabular}
\end{ruledtabular}
\end{table}

Among the ten phases, (II) and (IV) are trivial insulators. Phases (V), (VIII), (IX), and (X) are strong TIs. Phase (VII) is a HOTI and also a weak TI. Phases (I) and (III) are HOTIs with a trivial weak TI index. Although (I) and (III) share the same topological indices, they are not the same: Table~\ref{table1} shows that (I) and (III) correspond to band inversions at $\Gamma$ and at $R$, respectively.

In Sec.~\ref{mat}, we argue using the Topological Materials Database \cite{bradlyn2017topological,vergniory2019complete} that Ca$_3$SnO, Ca$_3$PbO, Ca$_3$GeO, Ba$_3$PbO, Sr$_3$PbO, and Sr$_3$SnO reside in phase (I). 
Hence, in the remainder of the manuscript, we will mainly focus on the HOTIs in region (I), which are characterized by a double band inversion at $\Gamma$.

\section{\label{mc} Mirror Chern states}

We would like to numerically verify the nontrivial HOTI phases computed in the previous section by showing that the bulk and surfaces are gapped, but that there are gapless modes on certain ``hinges'' where two surfaces meet.
However, there is a crucial problem: the $z$-normal surface is not gapped, due to the nonzero mirror Chern numbers reported in Ref.~\cite{hsieh2014topological}.
(Other mirror-preserving surfaces may also display mirror Chern states, including Type-II Dirac cones \cite{Chiu2017typeII}.)
In this section, we first review the mirror Chern number and then discuss several schemes for gapping the mirror Chern surface states.
Finally, in Sec.~\ref{hinge}, we will present evidence of the gapless hinge states that appear after gapping the surfaces.

\subsection{Mirror Chern number}

There are two inequivalent mirror symmetries in the space group $Pm\bar{3}m$: $M_{z}$ and $M_{xy}$.
Each mirror symmetry leaves two planes in the Brillouin zone invariant; for example, the $k_z=0$ and $k_z=\pi$ planes are invariant under $M_{z}$.

Within a mirror-invariant plane, the Hamiltonian can be decomposed into two subspaces ${\cal H}_{\pm {i}}$, where the subscript $\pm i$ indicates the mirror eigenvalue (mirror squares to $-1$ in a system with spin-orbit coupling). For each subspace, one can calculate the Chern number $C_\pm$ for the occupied bands. The mirror Chern number of the mirror-invariant plane is defined as $C_m =( C_+ - C_- )/2$ \cite{Teo2008surface}.
In a time reversal invariant system, $C_+ = -C_-$, and, consequently, $C_m = C_+$.

Hsieh, et al \cite{hsieh2014topological}, first identified the topological nature of antiperovskites by pointing out that those with an inverted band structure have a nontrivial mirror Chern number (previous work had already identified that the double band inversion led to a trivial $\mathbb{Z}_2$ index \cite{Sun2010,Kariyado2012}.)
Ref.~\cite{hsieh2014topological} identified the topological phase with a linear $k\cdot p$ model, which can be obtained by expanding (\ref{Hameqn}) to linear order. It is useful to introduce three new parameters:
\begin{equation}
\label{eqn_v}
 v_d = (v_1-2v_2)/2,~~ v_s = (2v_1+v_2)/2
\end{equation}
\begin{equation}
\label{eqn_R}
    R = v_d/v_s
\end{equation}
Ref.~\cite{hsieh2014topological} reported a phase transition when $|R|=1$, where the bulk gap closes, allowing the mirror Chern numbers to change.


By analytically and numerically~\cite{pythtb} studying the low-energy surface spectrum, we verify the results of Ref.~\cite{hsieh2014topological} in phase (I).
As mentioned at the end of Sec.~\ref{index},  this is the regime that describes the topologically nontrivial antiperovskite compounds.
A detailed analysis is in Appendix~\ref{surface_theory}; in summary, since the only band inversion in phase (I) occurs at the $\Gamma$ point, the $k_z=\pi$ mirror invariant plane is trivial, while the mirror Chern numbers in the $k_z=0$ ($M_z$-invariant) and $k_x=-k_y$ ($M_{xy}$-invariant) planes are enumerated in Table \ref{table3}, which agrees with a similar table in Ref.~\cite{PhysRevX.8.041026}. 
The phase transitions require the bulk band gap to close, which occurs at $R=0,~\pm 1$, as discussed in \cite{hsieh2014topological}. Further calculations (see Appendix \ref{surface_theory}) also reveal Lifshitz transitions for the gapless surface states at $R=\pm1/2$, which go beyond the analysis in \cite{hsieh2014topological}. When $|R|<1/2$, a gapless surface state crosses the Fermi surface only once, while when $1/2<|R|<1$, the state crosses the Fermi surface three times. \par

\begin{table}[b]
\caption{\label{table3} Table of mirror Chern number for $M_z$ and $M_{xy}$ mirror invariant planes. This result is only valid for phase (I) where the band inversion occurs at $\Gamma$.}
\begin{tabular}{|c|c|c|c|c|}
\hline
$R$ & $(-\infty,-1)$ & $~(-1,0)~$ & $~~(0,1)~~$ & $~(1,\infty)~$\\
\hline
$~C_m(M_z)~$ & -2 & 2 & 2 & -2\\
\hline
$~C_m(M_{xy})~$ & 0 & 2 & -2 & 0\\
\hline
\end{tabular}
\end{table}

\begin{figure}
\centering
\subfigure[][]{%
\label{bulksurface-a}%
\includegraphics[width=4.1cm]{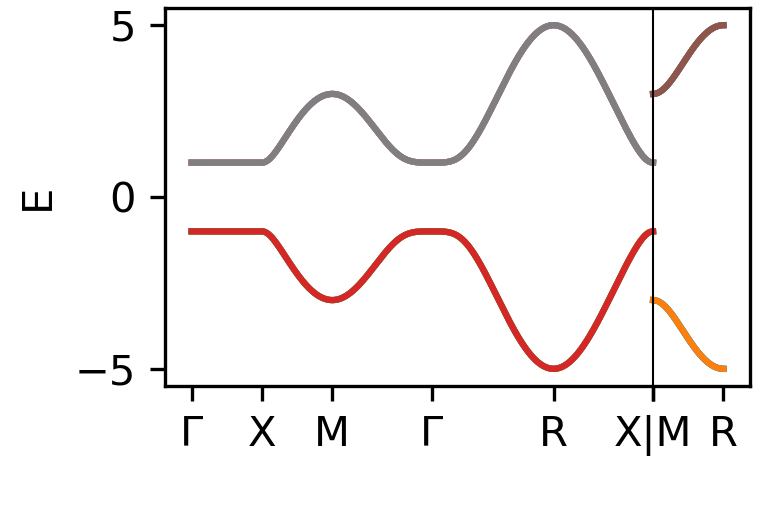}}%
\hspace{8pt}%
\subfigure[][]{%
\label{bulksurface-b}%
\includegraphics[width=4.1cm]{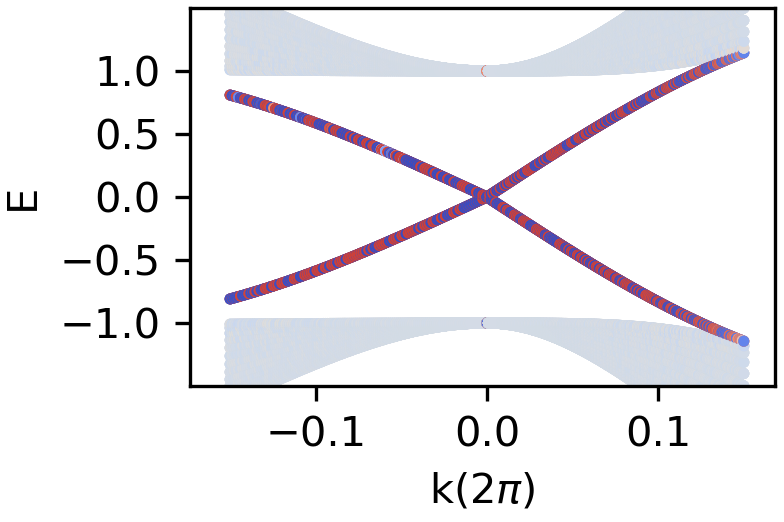}} 
\caption{Energy spectrum of the Hamiltonian (\ref{Hameqn}) in phase (I). \subref{bulksurface-a} Bulk states. Every band is four-fold degenerate in the special case $v_s=0$. \subref{bulksurface-b} Surface states on the $\hat{z}$-normal surface. Each line is four-fold degenerate, where one two-fold degeneracy comes from two surfaces of the slab, and the other from setting $v_s=0$. 
The horizontal axis travels along two mirror-invariant lines in the surface Brillouin zone, starting along the $\bar{X}-\bar{\Gamma}$ direction ($[-0.1,0]$) and then turning to the $\bar{\Gamma}-\bar{M}$ direction ($[0,0.1]$).
The parameters in (\ref{Hameqn}) are: $m=-1$, $\alpha_1=-\alpha_2 = 2$, $\beta_{1,2}=\gamma_{1,2}=0$, $v_d = 1$, $v_s = 0$.  
  }%
\label{bulksurface}%
\end{figure}

Fig.~\ref{bulksurface} shows the bulk bands and mirror Chern surface states for parameter values $m=-1$, $\alpha_1=-\alpha_2 = 2$, $\beta_{1,2}=\gamma_{1,2}=0$, $v_d = 1$, $v_s = 0$. The parameters are chosen to make the system reside in phase (I) with $\kappa=2$, $\nu_w=0$, and $C_m(M_z)=-2$, $C_m(M_{xy})=0$ with $R=\infty$.
Since $v_s=0$, every bulk and surface state has a two-fold degeneracy, in addition to the two-fold degeneracy from ${\cal P T}$ symmetry; combined, all states are four-fold degenerate, as shown in Fig.~\ref{bulksurface-a}.
Fig.~\ref{bulksurface-b} shows that along the $\bar{\Gamma}-\bar{M}$ direction (overbars denote high-symmetry points in the surface Brillouin zone), the surface states are gapless, despite the fact that $C_m(M_{xy})=0$, because setting $\beta_{1,2}=0$ enlarges the symmetries of the Hamiltonian. Nonetheless, the remainder of this section will continue using this choice of parameters unless stated otherwise; since we seek to gap the mirror surface states, the extra symmetries will soon be broken.\par

\subsection{\label{gap}Gapping the mirror Chern surface states}

As mentioned at the start of this section, the $\mathbb{Z}_4$-indicated hinge states will be obscured by the presence of the mirror Chern surface states.
A transport measurement to detect the hinge modes would also be dominated by the mirror Chern surface states. 
Thus, in order to observe the hinge states, we need to gap the mirror surface states.
We now propose three mechanisms to gap the mirror Chern surface states; similar ideas were pursued in Ref.~\cite{schindler2018higher} to reveal the hinge states in other types of higher-order topological insulators with mirror Chern surface states.\par

\begin{figure}
\centering
\subfigure[][]{%
\label{cleave-a}%
\includegraphics[width = 3cm]{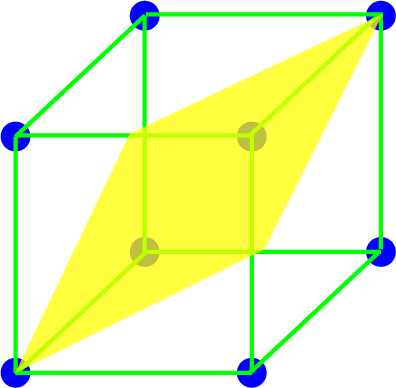}}%
\hspace{8pt}%
\subfigure[][]{%
\label{cleave-b}%
\includegraphics[width = 5cm]{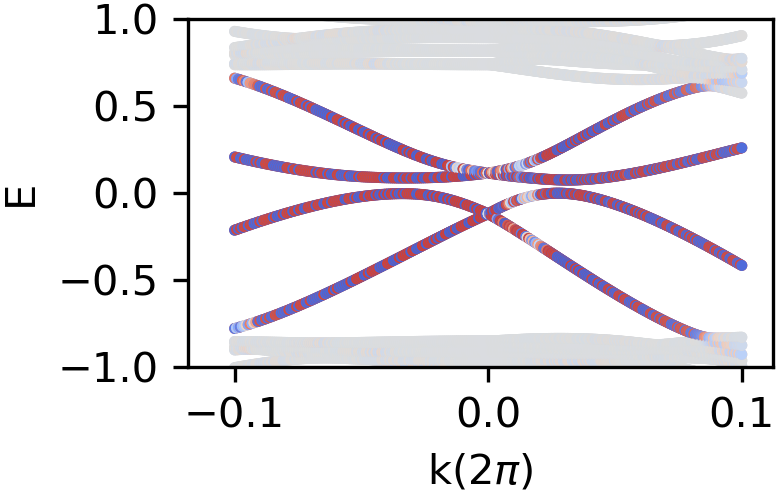}} 
\caption{Slab with a mirror-breaking normal. \subref{cleave-a} The yellow plane has normal vector $a_1 = (-2,1,1)$. Cleaving the crystal with an $a_1$-normal surface will break all mirror symmetries.
\subref{cleave-b} Surface states for a slab with the $a_1$-normal surface: a small gap opens due to the broken mirror symmetry. In this calculation, we have chosen a different set of parameters to get a relatively larger gap: $m=-1$, $\alpha_1=-\alpha_2 = 5$, $\beta_1=\beta_2=2$, $\gamma_{1,2}=0$, $v_d = 1$, $v_s = 2$. Horizontal axis is momentum, traversing a path along $\bar{X}-\bar{\Gamma}-\bar{M}$. 
}%
\label{cleave}%
\end{figure}

The first idea is to cleave the lattice such that the resulting surface breaks all the mirror symmetries. An example of a low-symmetry surface is shown in Fig.~\ref{cleave-a} and the corresponding
surface states are calculated in Fig.~\ref{cleave-b}. A small surface gap is opened along both the $\bar{X}-\bar{\Gamma}$ and $\bar{\Gamma}-\bar{M}$ directions. In this calculation, we have chosen a different set of parameters than in the previous section to get a relatively larger gap.
Specifically, the quadratic $\beta$ terms are set to be non-zero and break particle-hole symmetry, so that the four-fold degeneracy at $\Gamma$ is split into two two-fold degenerate bands. The details of this argument are in Appendix~\ref{surface_theory}. The $\alpha$ terms are then chosen so that the system remains in phase (I). Since $R=1/2$, one of the mirror Chern surface bands is almost flat at $\Gamma$.

We find that the surface gap opened by a mirror-breaking surface is usually very small and our model does not contain a mechanism to tune it. 
However, in a real sample, the surface gap will depend strongly on the microscopic details of the surface dangling bonds and their interaction with the environment.
In particular, growth of a capping layer on the surface or adsorption of organic molecules could dramatically change the surface electronic structure.

\begin{figure}
\centering
\subfigure[][]{%
\label{sp-a}%
\includegraphics[width=4.1cm]{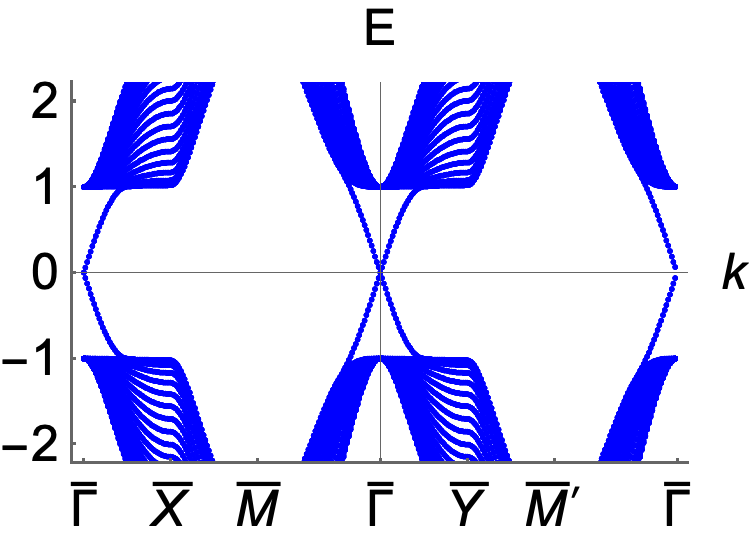}}%
\hspace{8pt}%
\subfigure[][]{%
\label{sp-b}%
\includegraphics[width=4.1cm]{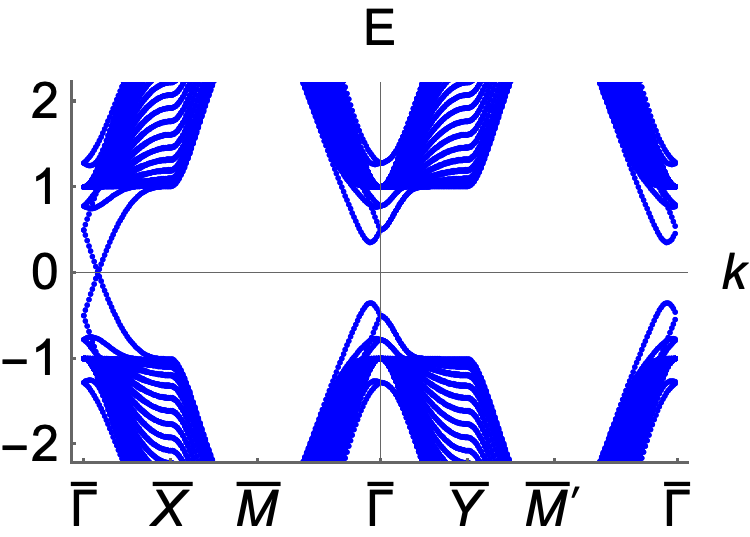}} 
\subfigure[][]{%
\label{sp-c}%
\includegraphics[width=4.1cm]{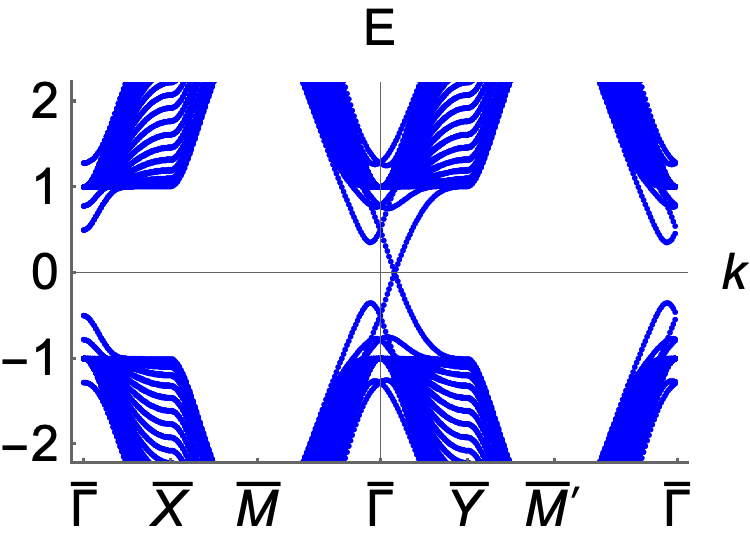}}%
\hspace{8pt}%
\subfigure[][]{%
\label{sp-d}%
\includegraphics[width=4.1cm]{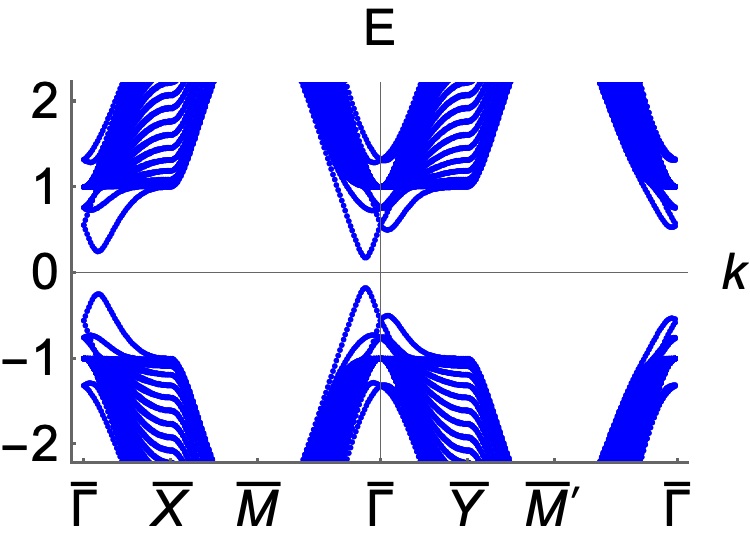}}
\caption{
Mirror Chern surface states can be moved or gapped by tuning the surface potential, $(V_1,V_2)$. The surface potentials are set as $V_1 = \delta_{a1}=-\delta_{a2}$, $V_2 = \delta_{b1}=-\delta_{b2}$. \subref{sp-a} Gapless mirror surface states without the surface potential; \subref{sp-b} with the surface potential $(0.5, 0)$, the gapless states are moved along $\bar{\Gamma} -\bar{X}$ and gapped along other directions; \subref{sp-c} with the surface potential $(0, 0.5)$, the gapless states are moved along $\bar{\Gamma} -\bar{Y}$ and gapped along other directions; \subref{sp-d} with the surface potential $(0.5, 0.25)$, the mirror surface states are completely gapped.
}%
\label{figsp}%
\end{figure}

This is related to our second mechanism to gap the mirror Chern surface states, which is to consider a mirror-breaking surface perturbation on a surface that would otherwise preserve mirror symmetry.
This could be accomplished by growing a thin film on both the top and bottom surfaces, such that inversion and time reversal are preserved, but the mirror symmetries are broken.
Then the boundary states will be derived from the bulk Hamiltonian plus a mirror-breaking surface potential. The lowest order ($k$-independent) surface potential that satisfies the symmetry requirements is:
\begin{align}
\label{eqn_surface_potential}
    V_{surface} =  \begin{bmatrix}
    V(\delta_{a1},\delta_{b1},\delta_{c1},\delta_{d1}) &
   0 \\
   0&
  V(\delta_{a2},\delta_{b2},\delta_{c2},\delta_{d2})
   \end{bmatrix},
\end{align}
where
\begin{equation}
   V(\delta_a,\delta_b,\delta_c,\delta_d) =  \delta_a \sigma_x\tau_z - \delta_b \sigma_y\tau_z+\delta_c\sigma_0\tau_x-\delta_d\sigma_0\tau_y
\end{equation}

Table~\ref{table4} presents the nine mirror symmetries and which terms in Eq.~(\ref{eqn_surface_potential}) break each one. For example, $M_{xy}$ would be broken when $\delta_c$ or $\delta_a-\delta_b$ is nonzero. \par

\begin{table}[b]
\caption{\label{table4} Table of mirror symmetries and their symmetry breaking k-independent surface potentials. The special cases shown in Fig.~\ref{figsp} (b)(c)(d) are also listed in the three columns labeled $\delta_{a}$, $\delta_{b}$, and $\delta_{a}=2\delta_{b}$. X indicates symmetry is broken, while \checkmark indicates symmetry is still preserved.}
\begin{tabular}{|c|c|c|c|c|}
\hline
 mirror & broken by & $\quad~\delta_{a}~\quad$ & $\quad~\delta_{b}~\quad$ & $\delta_{a}=2\delta_{b}$\\
 \hline
 $M_x$ & $\delta_{a}$, $\delta_{d}$ &X  &\checkmark &X\\
 $M_y$ & $\delta_{b}$, $\delta_{d}$ &\checkmark &X & X\\
 $M_z$ & $\delta_{a}$, $\delta_{b}$ & X&X &X \\
 $M_{xy}$ & $~\delta_{c}$, $\delta_{a}-\delta_b~$ &X & X &X\\
 $M_{\bar{x}y}$ & $~\delta_{c}$, $\delta_{a}+\delta_b~$ &X &X &X \\
 $M_{yz}$ & $~\delta_{c}$, $\delta_{a}-\delta_d~$ &X &\checkmark & X\\
 $M_{\bar{y}z}$ & $~\delta_{c}$, $\delta_{a}+\delta_d~$ &X &\checkmark & X\\
 $M_{zx}$ & $~\delta_{c}$, $\delta_{b}+\delta_d~$ &\checkmark & X&X\\
 $M_{\bar{z}x}$ & $~\delta_{c}$, $\delta_{b}-\delta_d~$ &\checkmark & X&X\\
\hline
\end{tabular}
\end{table}

Fig.~\ref{figsp} shows the surface states calculated with a $\hat{z}$-normal surface on which different surface potentials have been added to the top and bottom layers (so that inversion symmetry is preserved). Since the surface potentials break mirror symmetries, the $C_4$ symmetry in the surface plane is also broken. So we present a band structure along the path $\bar{\Gamma}-\bar{X}-\bar{M}-\bar{\Gamma}-\bar{Y}-\bar{M}'-\bar{\Gamma}$ in the surface Brillouin zone. \par

We can understand which paths become gapped or remain gapless by studying which mirror symmetries are broken or preserved. For a slab with a $\hat{z}$-normal, $M_{yz}$, $M_{\bar{y}z}$, $M_{zx}$ and $M_{\bar{z}x}$ have been broken by the geometry; $M_z$ is preserved in the bulk, but broken on a single surface (it relates the two surfaces to each other). Thus, only $M_x$, $M_y$, $M_{xy}$ and $M_{\bar{x}y}$ mirror surface states can appear on the slab surface. In the case of Fig.~\ref{sp-b}, $M_y$ is preserved, so there is a crossing along $\bar{\Gamma}-\bar{X}$ (surface projection of the $k_y=0$ plane). In the case of Fig.~\ref{sp-c}, $M_{x}$ is preserved, so there is a crossing along $\bar{\Gamma}-\bar{Y}$ (surface projection of the $k_x=0$ plane). In the case of Fig.~\ref{sp-d}, all mirror symmetries are broken, thus the surface states are gapped everywhere.\par

\begin{figure}
\centering
\subfigure[][]{%
\label{strain-a}%
\includegraphics[width=4.1cm]{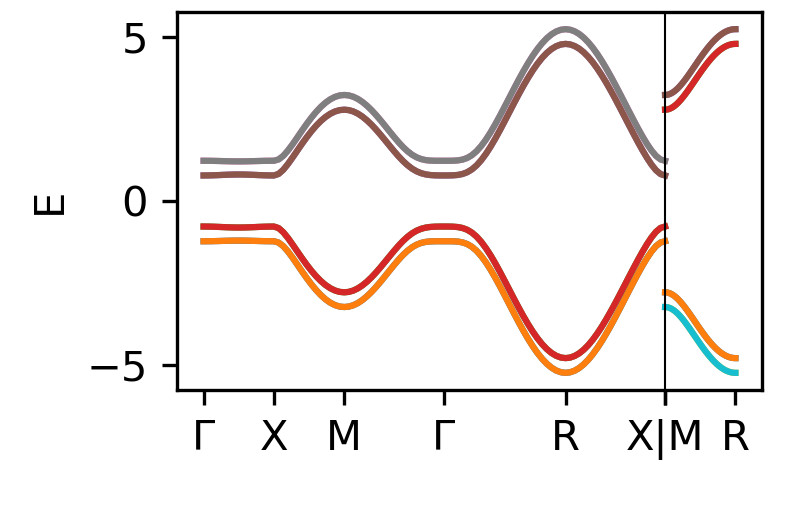}}%
\hspace{8pt}%
\subfigure[][]{%
\label{strain-b}%
\includegraphics[width=4.1cm]{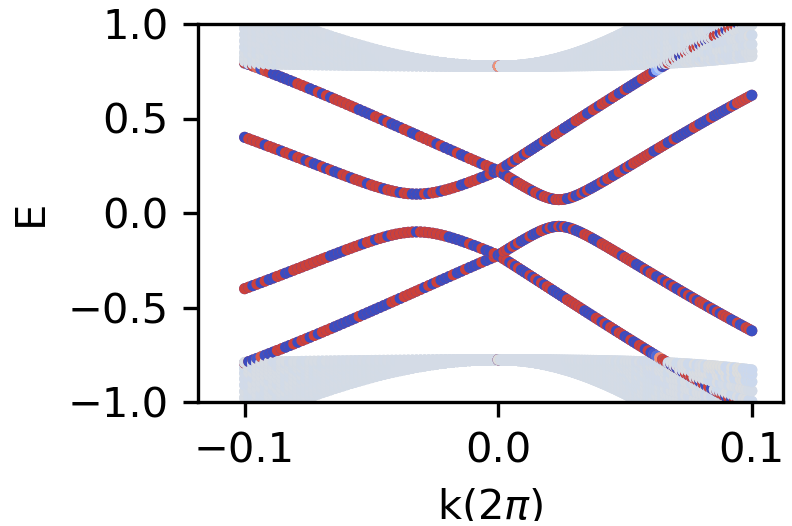}} 
\caption{
Energy spectrum with strain parameters $V_1= \delta_{a1}=-\delta_{a_2}=0.1$, $V_2=\delta_{b1}=-\delta_{b_2}=0.05$, which \subref{strain-a} shifts the bulk bands and \subref{strain-b} opens the surface gap. If the strain energy $V_1$ and $V_2$ is too large (more than 0.5), the bulk will cease to be an insulator.
}%
\label{figstrain}%
\end{figure}

The third way to gap the mirror Chern surface states is to apply strain to the lattice to break the mirror symmetry. The strain term takes the same form as the surface potential (\ref{eqn_surface_potential}). But there is a small difference. While in a heterostructure, the surface potential only appears in the first several layers near the boundary, for the strained system, the strain term appears at every layer inside the bulk.
We would expect the strain to result from an external force, although it could perhaps be engineered by the growth conditions.
Fig.~\ref{strain-a} shows the bulk states shifted by the strain term. Fig.~\ref{strain-b} shows the surface gap opened by strain.
\par

\section{Hinge states\label{hinge}}

Armed with three mechanisms for gapping the mirror Chern surface states, we numerically demonstrate the presence of the higher-order hinge states in the presence of a non-zero surface potential or strain (\ref{eqn_surface_potential}).

We consider a rod geometry with a square cross section that is finite in the $x$ and $y$ directions and infinite in the $z$ direction; thus, we plot the energy spectrum as a function of $k_z$, using the same parameters as in Sec. \ref{gap}.
\par

\begin{figure}
\centering
\subfigure[][]{%
\label{hingefig-a}%
\includegraphics[width=4.5cm]{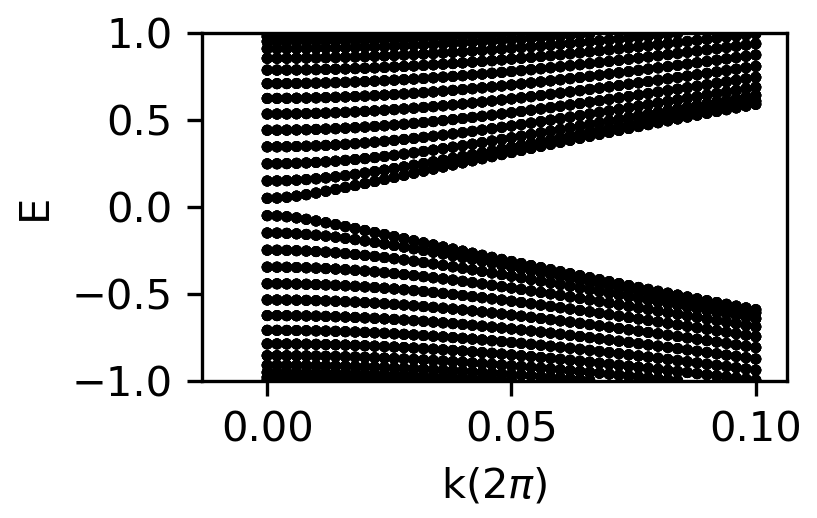}}%
\hspace{8pt}%
\subfigure[][]{%
\label{hingefig-b}%
\includegraphics[width=3.5cm]{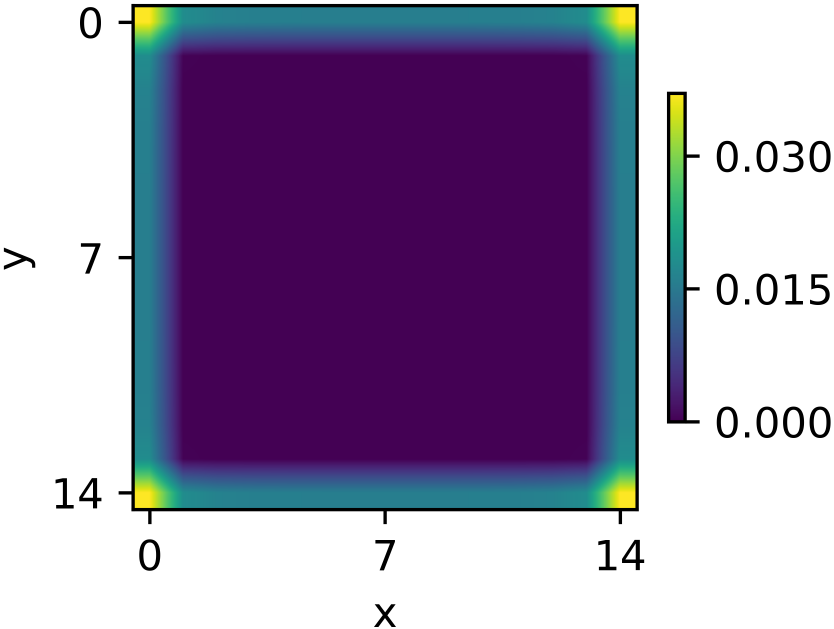}} \\
\subfigure[][]{%
\label{hingefig-c}%
\includegraphics[width=4.5cm]{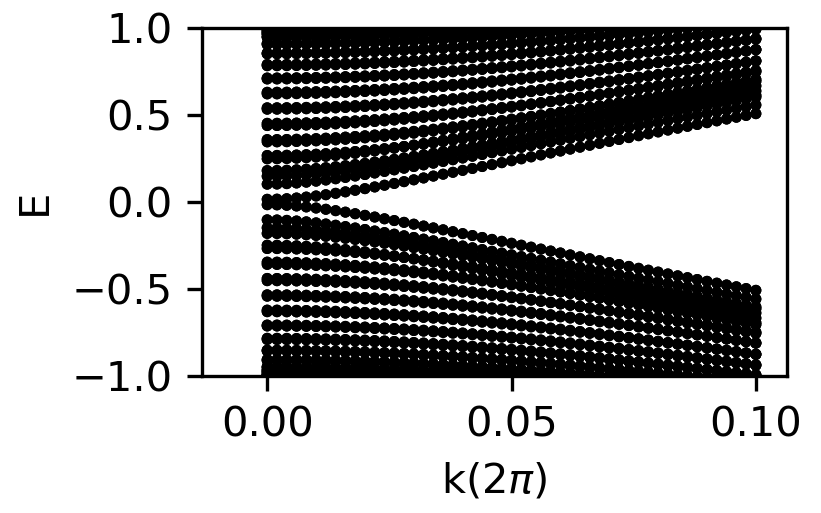}}%
\hspace{8pt}%
\subfigure[][]{%
\label{hingefig-d}%
\includegraphics[width=3.5cm]{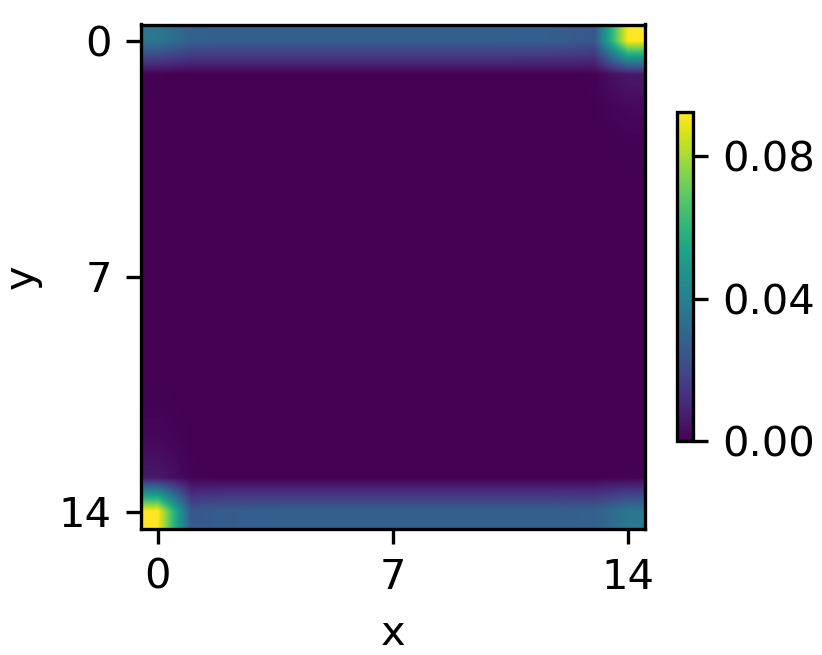}}\\
\subfigure[][]{%
\label{hingefig-e}%
\includegraphics[width=4.5cm]{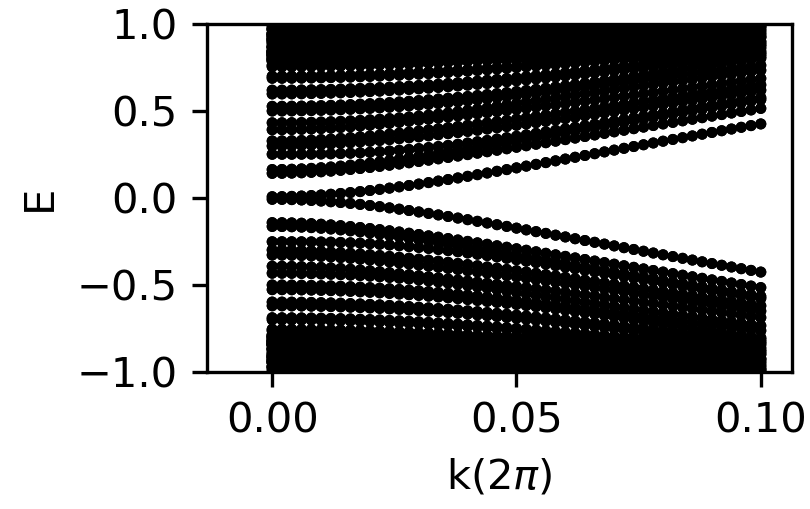}}%
\hspace{8pt}%
\subfigure[][]{%
\label{hingefig-f}%
\includegraphics[width=3.5cm]{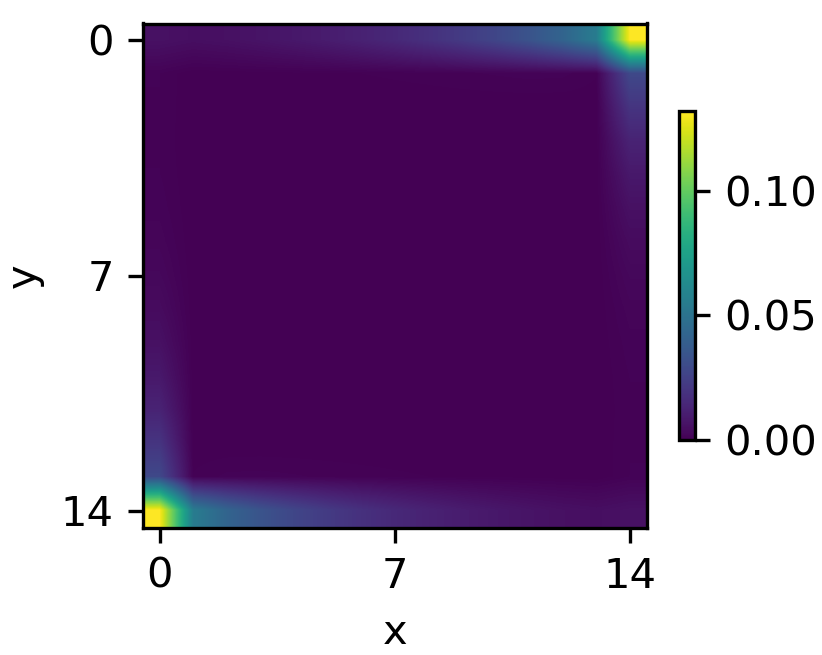}}
\caption{
Hinge states with surface potentials. \subref{hingefig-a}  Energy spectrum of a rod without a surface potential: gapless states come from both surface states and hinge states; \subref{hingefig-b} position space distribution for one of the wavefunctions with zero energy: the wavefunction has a large weight at the four corners, but there is a non-negligible weight on all surfaces, which prevents the hinge states from being exponentially localized.
\subref{hingefig-c} Energy spectrum of a rod with surface potential $V_1 = 0.1$, $V_2 = 0.05$: gapless helical hinge states appear; \subref{hingefig-d}  position space distribution for one of the wavefunctions with zero energy: the wavefunction now has negligible weight on the left and right surfaces. \subref{hingefig-e}  Energy spectrum of a rod with bulk strain given by: 
$V_1 = 0.1$, $V_2 = 0.05$
gapless helical hinge states appear in the surface gap; \subref{hingefig-f} the position space distribution of a wavefunction with zero energy reveals localization in the top-right and bottom-left corners. In all cases, the rod has 15 sites in the $x$ and $y$ directions and is infinite in the $z$ direction. 
}%
\label{hingefig}%
\end{figure}

The results are shown in Fig.~\ref{hingefig}. Figures~\subref{hingefig-a}, \subref{hingefig-c}, and \subref{hingefig-e} show the dispersion in momentum space, while~\subref{hingefig-b}, \subref{hingefig-d}, and \subref{hingefig-f} show the weight distribution in real space for a specific state at Fermi level (the other zero energy states have the same weight distribution). ~\subref{hingefig-a} and \subref{hingefig-b} do not have any surface potential or strain. Thus, all mirror symmetries are preserved and the zero energy states have weight on the surfaces as well as the hinges in \subref{hingefig-b}; in addition, the energy spectrum in \subref{hingefig-a} is characteristic of a gapless surface. ~\subref{hingefig-c} and \subref{hingefig-d} are plotted with a surface potential added to the left and right edges. ~\subref{hingefig-c} shows zero-energy states in a small surface gap. Since the mirror symmetries are all broken by the surface potential at the left and right surfaces, there is no weight on those two edges in \subref{hingefig-d}. 
~\subref{hingefig-e} and \subref{hingefig-f} are plotted with bulk strain. All mirror symmetries are broken by the strain term in bulk, and as a result, the surface gap is opened (approximately $|E| < .2$). The strained hinge states are localized at only two corners in real space, as shown in ~\subref{hingefig-f}. 
\par
{The real space distribution is localized at the left-bottom and right-top corners in Fig.~\ref{hingefig-f}. However, for different strain parameters, the hinge states are localized at the left-top and right-bottom corners. 
This behavior can be understood by realizing that the antiperovskites also have a higher-order invariant protected by even mirror Chern number, similar to SnTe \cite{schindler2018higher}. 
We discuss this connection and the phase transition between the two configurations of hinge states in Appendix.~\ref{appc}
}

\section{Materials: Antiperovskites\label{mat}}

The motivation for establishing and analyzing our tight-binding model (\ref{Hameqn}) was to determine all possible topological phases consistent with the symmetry and orbital content described in Sec.~\ref{Ham}; we summarized the analysis with the phase diagram in Fig.~\ref{pdfigure}.

\begin{table}[b]
\caption{\label{table5}Table of antiperovskites with $\mathbb{Z}_4$ index $\kappa=2$. Number of occupied bands with positive/negative parity at TRIMs is listed. The last column shows the smallest direct gaps (DG) for these materials. }
\begin{ruledtabular}
\begin{tabular}{cccccccc}
$A_3BX$ & $\kappa$ & $\nu_w$ &     \begin{tabular}{@{}c@{}}$\Gamma$ \\ $n_+\quad n_-$\end{tabular}&     \begin{tabular}{@{}c@{}}$X$ \\ $n_+\quad n_-$\end{tabular}&     \begin{tabular}{@{}c@{}}$M$ \\ $n_+\quad n_-$\end{tabular}&     \begin{tabular}{@{}c@{}}$R$ \\ $n_+\quad n_-$\end{tabular} &DG(meV) 
\\
\hline
Ca$_3$SnO & 2 &0 & 8 \quad 8 &8 \quad 8 &4 \quad 12 &8 \quad 8 & 18.1 \\
Ca$_3$PbO & 2 &0 & 8 \quad 8 &8 \quad 8 &4 \quad 12 &8 \quad 8 &44.5 \\
Ca$_3$GeO & 2 &0 & 8 \quad 8 &8 \quad 8 &4 \quad 12 &8 \quad 8 &3.1 \\
Ba$_3$PbO & 2 &0 & 14 ~ 26&22 \quad 18&18 \quad 22 &14 \quad 26 &5.9 \\
Sr$_3$PbO & 2 &0 & 14 ~ 26&22 \quad 18&18 \quad 22 &14 \quad 26 &54.9 \\
Sr$_3$SnO & 2 &0 & 14 ~ 26&22 \quad 18&18 \quad 22 &14 \quad 26 &52.2 \\
Sr$_3$BiN & 2 &0 & 14 ~ 26&22 \quad 18&18 \quad 22 &14 \quad 26 &8.6 \\
Ti$_3$TlN & 2 &1 &14 \quad 6  &10 \quad 10& ~4 \quad 16 &12 \quad 8 &9.3 \\
\end{tabular}
\end{ruledtabular}
\end{table}

We now seek to place known antiperovskites into the phase diagram in Fig.~\ref{pdfigure}. This can be readily accomplished by using the Topological Materials Database \cite{bradlyn2017topological,vergniory2019complete}, which lists the topological indices for each reported compound.
Furthermore, the phases with the same topological indices (i.e., (I) and (III), (II) and (IV), and (V) and (X)) can be distinguished because the database also lists the inversion eigenvalues at each TRIM point.


We find that eight reported antiperovskite materials, listed in Table~\ref{table5}, have a nontrivial HOTI index $\kappa$, and further, that six of these -- Ca$_3$SnO, Ca$_3$PbO, Ca$_3$GeO, Ba$_3$PbO, Sr$_3$PbO, and Sr$_3$SnO -- are in phase (I). 
They can be understood as four occupied bands with a double band inversion at $\Gamma$, plus two trivial bands of parity $-1$, and several additional bands at lower energies that are separated from the six bands by a large gap. This justifies our analysis of this phase in later sections. Since many of these materials are easily cleavable, we propose that they are a good place to search for gapless modes on hinges or possibly on step edges \cite{queiroz2019partial}.
There may be additional HOTIs realized in the $f$-electron antiperovskites (for example, compounds studied in \cite{PhysRevB.99.205126}), which are not in the Topological Materials Database \cite{bradlyn2017topological,vergniory2019complete}.

The size of the bulk band gap is determined by the spin orbit coupling strength, which gaps the Dirac cone along $\Gamma-X$, as shown in Fig.~\ref{figcrystal}(b).
Three compounds in Table~\ref{table5} have gaps {$\Delta \sim 50$} meV, which is large enough to be resolved in experiment. 
{However, the ability to resolve the hinge states depends not only on the bulk band gap, but also on the surface band gap, which 
is determined by the mirror-symmetry breaking mechanism that gaps the mirror Chern surface states (see Sec.~\ref{gap}.)}


{The gap at $\Gamma$, $\Delta_\Gamma$ in Fig.~\ref{figcrystal}(b), is much larger than the bulk band gap; specifically, $\Delta_\Gamma \gtrsim 10\Delta$.
Therefore, even in materials with a small gap, the hinge states may be visible in a momentum-resolved measurement by properly choosing the surface direction so that the smallest gap $\Delta$, along $\Gamma-X$, does not project onto $\Gamma$.
For example, the surfaces normal to the $(111)$ and $(\bar{1}11)$ directions fulfill this condition. In this case, the gap $\Delta_1$ along $\Gamma-R$ is projected onto $\Gamma$, but it is much larger than $\Delta$.}


\section{Outlook}
\label{outlook}

In this manuscript, we have shown that antiperovskites with an inverted band structure fall into the newly discovered HOTI phase, protected by time reversal and inversion symmetry.
These HOTIs will display gapless helical modes on hinges where two surfaces meet; such hinge modes are similar to those on the edge of a two-dimensional topological insulator and thus present another route to finding Majorana fermions in materials with a trivial $\mathbb{Z}_2$ index, when combined with superconductivity \cite{queiroz2019splitting,fu2008superconducting,nilsson2008splitting,fu2009josephson,tanaka2009manipulation,linder2010unconventional};
in fact, superconductivity has been observed in one of our candidate HOTIs, doped Sr$_3$SnO \cite{Oudah_2016}.
Helical modes may also be observed along the edges of thin films, which have recently been grown for Sr$_3$SnO \cite{ma2019realization}.
Finally, another route to observing the gapless helical modes may be to perform STM on crystal defects: future calculations are necessary to determine whether dislocations or step edges host gapless helical modes, as has been demonstrated for other HOTIs \cite{queiroz2019partial}.

Importantly, many antiperovskites are stable and experimentally well studied.
In particular, the compounds Ca$_3$PbO, Ca$_3$SnO, Sr$_3$PbO, and Sr$_3$SnO have been the subject of several recent experiments \cite{Suetsugu_2018,Oudah_2016,PhysRevB.96.155109,Kitagawa_2018,ma2019realization,Rost2019inverse,Huang2019unusual}. 
Thus, our results present a promising avenue to pursue experimental studies of HOTIs, for which very few other candidates have been identified.
We hope that our work motivates an experimental search for hinge modes in these materials, as well as ab initio studies on the effects of strain and surface perturbations in order to determine a realistic estimate of the surface gap.

\begin{acknowledgments}
The authors thank Barry Bradlyn and Cyrus Dreyer for useful conversations on surface terminations and Alec Wills for computational assistance.
The authors also thank Stony Brook Research Computing and Cyberinfrastructure and the Institute for Advanced Computational Science at Stony Brook University for access to the SeaWulf computing system, which was made possible by  a \$1.4M National Science Foundation grant (\#1531492).
J.C. acknowledges support from the Flatiron Institute, a division of the Simons Foundation.
\end{acknowledgments}


\bibliography{Antiperovskite.bib}

\begin{thebibliography}{53}%
\makeatletter
\providecommand \@ifxundefined [1]{%
 \@ifx{#1\undefined}
}%
\providecommand \@ifnum [1]{%
 \ifnum #1\expandafter \@firstoftwo
 \else \expandafter \@secondoftwo
 \fi
}%
\providecommand \@ifx [1]{%
 \ifx #1\expandafter \@firstoftwo
 \else \expandafter \@secondoftwo
 \fi
}%
\providecommand \natexlab [1]{#1}%
\providecommand \enquote  [1]{``#1''}%
\providecommand \bibnamefont  [1]{#1}%
\providecommand \bibfnamefont [1]{#1}%
\providecommand \citenamefont [1]{#1}%
\providecommand \href@noop [0]{\@secondoftwo}%
\providecommand \href [0]{\begingroup \@sanitize@url \@href}%
\providecommand \@href[1]{\@@startlink{#1}\@@href}%
\providecommand \@@href[1]{\endgroup#1\@@endlink}%
\providecommand \@sanitize@url [0]{\catcode `\\12\catcode `\$12\catcode
  `\&12\catcode `\#12\catcode `\^12\catcode `\_12\catcode `\%12\relax}%
\providecommand \@@startlink[1]{}%
\providecommand \@@endlink[0]{}%
\providecommand \url  [0]{\begingroup\@sanitize@url \@url }%
\providecommand \@url [1]{\endgroup\@href {#1}{\urlprefix }}%
\providecommand \urlprefix  [0]{URL }%
\providecommand \Eprint [0]{\href }%
\providecommand \doibase [0]{https://doi.org/}%
\providecommand \selectlanguage [0]{\@gobble}%
\providecommand \bibinfo  [0]{\@secondoftwo}%
\providecommand \bibfield  [0]{\@secondoftwo}%
\providecommand \translation [1]{[#1]}%
\providecommand \BibitemOpen [0]{}%
\providecommand \bibitemStop [0]{}%
\providecommand \bibitemNoStop [0]{.\EOS\space}%
\providecommand \EOS [0]{\spacefactor3000\relax}%
\providecommand \BibitemShut  [1]{\csname bibitem#1\endcsname}%
\let\auto@bib@innerbib\@empty
\bibitem [{\citenamefont {Bradlyn}\ \emph {et~al.}(2017)\citenamefont
  {Bradlyn}, \citenamefont {Elcoro}, \citenamefont {Cano}, \citenamefont
  {Vergniory}, \citenamefont {Wang}, \citenamefont {Felser}, \citenamefont
  {Aroyo},\ and\ \citenamefont {Bernevig}}]{bradlyn2017topological}%
  \BibitemOpen
  \bibfield  {author} {\bibinfo {author} {\bibfnamefont {B.}~\bibnamefont
  {Bradlyn}}, \bibinfo {author} {\bibfnamefont {L.}~\bibnamefont {Elcoro}},
  \bibinfo {author} {\bibfnamefont {J.}~\bibnamefont {Cano}}, \bibinfo {author}
  {\bibfnamefont {M.}~\bibnamefont {Vergniory}}, \bibinfo {author}
  {\bibfnamefont {Z.}~\bibnamefont {Wang}}, \bibinfo {author} {\bibfnamefont
  {C.}~\bibnamefont {Felser}}, \bibinfo {author} {\bibfnamefont
  {M.}~\bibnamefont {Aroyo}},\ and\ \bibinfo {author} {\bibfnamefont {B.~A.}\
  \bibnamefont {Bernevig}},\ }\bibfield  {title} {\bibinfo {title} {Topological
  quantum chemistry},\ }\href@noop {} {\bibfield  {journal} {\bibinfo
  {journal} {Nature}\ }\textbf {\bibinfo {volume} {547}},\ \bibinfo {pages}
  {298} (\bibinfo {year} {2017})}\BibitemShut {NoStop}%
\bibitem [{\citenamefont {Po}\ \emph {et~al.}(2017)\citenamefont {Po},
  \citenamefont {Vishwanath},\ and\ \citenamefont {Watanabe}}]{Po2017}%
  \BibitemOpen
  \bibfield  {author} {\bibinfo {author} {\bibfnamefont {H.~C.}\ \bibnamefont
  {Po}}, \bibinfo {author} {\bibfnamefont {A.}~\bibnamefont {Vishwanath}},\
  and\ \bibinfo {author} {\bibfnamefont {H.}~\bibnamefont {Watanabe}},\
  }\bibfield  {title} {\bibinfo {title} {Symmetry-based indicators of band
  topology in the 230 space groups},\ }\href@noop {} {\bibfield  {journal}
  {\bibinfo  {journal} {Nature Communications}\ }\textbf {\bibinfo {volume}
  {8}},\ \bibinfo {pages} {50} (\bibinfo {year} {2017})}\BibitemShut {NoStop}%
\bibitem [{\citenamefont {Cano}\ \emph {et~al.}(2018)\citenamefont {Cano},
  \citenamefont {Bradlyn}, \citenamefont {Wang}, \citenamefont {Elcoro},
  \citenamefont {Vergniory}, \citenamefont {Felser}, \citenamefont {Aroyo},\
  and\ \citenamefont {Bernevig}}]{Cano2018building}%
  \BibitemOpen
  \bibfield  {author} {\bibinfo {author} {\bibfnamefont {J.}~\bibnamefont
  {Cano}}, \bibinfo {author} {\bibfnamefont {B.}~\bibnamefont {Bradlyn}},
  \bibinfo {author} {\bibfnamefont {Z.}~\bibnamefont {Wang}}, \bibinfo {author}
  {\bibfnamefont {L.}~\bibnamefont {Elcoro}}, \bibinfo {author} {\bibfnamefont
  {M.~G.}\ \bibnamefont {Vergniory}}, \bibinfo {author} {\bibfnamefont
  {C.}~\bibnamefont {Felser}}, \bibinfo {author} {\bibfnamefont {M.~I.}\
  \bibnamefont {Aroyo}},\ and\ \bibinfo {author} {\bibfnamefont {B.~A.}\
  \bibnamefont {Bernevig}},\ }\bibfield  {title} {\bibinfo {title} {Building
  blocks of topological quantum chemistry: Elementary band representations},\
  }\href {https://doi.org/10.1103/PhysRevB.97.035139} {\bibfield  {journal}
  {\bibinfo  {journal} {Phys. Rev. B}\ }\textbf {\bibinfo {volume} {97}},\
  \bibinfo {pages} {035139} (\bibinfo {year} {2018})},\ \Eprint
  {https://arxiv.org/abs/1709.01935} {arXiv:1709.01935 [cond-mat]} \BibitemShut
  {NoStop}%
\bibitem [{\citenamefont {Vergniory}\ \emph {et~al.}(2017)\citenamefont
  {Vergniory}, \citenamefont {Elcoro}, \citenamefont {Wang}, \citenamefont
  {Cano}, \citenamefont {Felser}, \citenamefont {Aroyo}, \citenamefont
  {Bernevig},\ and\ \citenamefont {Bradlyn}}]{Vergniory2017graph}%
  \BibitemOpen
  \bibfield  {author} {\bibinfo {author} {\bibfnamefont {M.~G.}\ \bibnamefont
  {Vergniory}}, \bibinfo {author} {\bibfnamefont {L.}~\bibnamefont {Elcoro}},
  \bibinfo {author} {\bibfnamefont {Z.}~\bibnamefont {Wang}}, \bibinfo {author}
  {\bibfnamefont {J.}~\bibnamefont {Cano}}, \bibinfo {author} {\bibfnamefont
  {C.}~\bibnamefont {Felser}}, \bibinfo {author} {\bibfnamefont {M.~I.}\
  \bibnamefont {Aroyo}}, \bibinfo {author} {\bibfnamefont {B.~A.}\ \bibnamefont
  {Bernevig}},\ and\ \bibinfo {author} {\bibfnamefont {B.}~\bibnamefont
  {Bradlyn}},\ }\bibfield  {title} {\bibinfo {title} {Graph theory data for
  topological quantum chemistry},\ }\href
  {https://doi.org/10.1103/PhysRevE.96.023310} {\bibfield  {journal} {\bibinfo
  {journal} {Phys. Rev. E}\ }\textbf {\bibinfo {volume} {96}},\ \bibinfo
  {pages} {023310} (\bibinfo {year} {2017})},\ \Eprint
  {https://arxiv.org/abs/1706.08529} {arXiv:1706.08529} \BibitemShut {NoStop}%
\bibitem [{\citenamefont {Bradlyn}\ \emph {et~al.}(2018)\citenamefont
  {Bradlyn}, \citenamefont {Elcoro}, \citenamefont {Vergniory}, \citenamefont
  {Cano}, \citenamefont {Wang}, \citenamefont {Felser}, \citenamefont {Aroyo},\
  and\ \citenamefont {Bernevig}}]{Bradlyn2018band}%
  \BibitemOpen
  \bibfield  {author} {\bibinfo {author} {\bibfnamefont {B.}~\bibnamefont
  {Bradlyn}}, \bibinfo {author} {\bibfnamefont {L.}~\bibnamefont {Elcoro}},
  \bibinfo {author} {\bibfnamefont {M.~G.}\ \bibnamefont {Vergniory}}, \bibinfo
  {author} {\bibfnamefont {J.}~\bibnamefont {Cano}}, \bibinfo {author}
  {\bibfnamefont {Z.}~\bibnamefont {Wang}}, \bibinfo {author} {\bibfnamefont
  {C.}~\bibnamefont {Felser}}, \bibinfo {author} {\bibfnamefont {M.~I.}\
  \bibnamefont {Aroyo}},\ and\ \bibinfo {author} {\bibfnamefont {B.~A.}\
  \bibnamefont {Bernevig}},\ }\bibfield  {title} {\bibinfo {title} {Band
  connectivity for topological quantum chemistry: Band structures as a graph
  theory problem},\ }\href {https://doi.org/10.1103/PhysRevB.97.035138}
  {\bibfield  {journal} {\bibinfo  {journal} {Phys. Rev. B}\ }\textbf {\bibinfo
  {volume} {97}},\ \bibinfo {pages} {035138} (\bibinfo {year} {2018})},\
  \Eprint {https://arxiv.org/abs/1709.01937} {arXiv:1709.01937} \BibitemShut
  {NoStop}%
\bibitem [{\citenamefont {Vergniory}\ \emph {et~al.}(2019)\citenamefont
  {Vergniory}, \citenamefont {Elcoro}, \citenamefont {Felser}, \citenamefont
  {Regnault}, \citenamefont {Bernevig},\ and\ \citenamefont
  {Wang}}]{vergniory2019complete}%
  \BibitemOpen
  \bibfield  {author} {\bibinfo {author} {\bibfnamefont {M.}~\bibnamefont
  {Vergniory}}, \bibinfo {author} {\bibfnamefont {L.}~\bibnamefont {Elcoro}},
  \bibinfo {author} {\bibfnamefont {C.}~\bibnamefont {Felser}}, \bibinfo
  {author} {\bibfnamefont {N.}~\bibnamefont {Regnault}}, \bibinfo {author}
  {\bibfnamefont {B.~A.}\ \bibnamefont {Bernevig}},\ and\ \bibinfo {author}
  {\bibfnamefont {Z.}~\bibnamefont {Wang}},\ }\bibfield  {title} {\bibinfo
  {title} {A complete catalogue of high-quality topological materials},\
  }\href@noop {} {\bibfield  {journal} {\bibinfo  {journal} {Nature}\ }\textbf
  {\bibinfo {volume} {566}},\ \bibinfo {pages} {480} (\bibinfo {year}
  {2019})}\BibitemShut {NoStop}%
\bibitem [{\citenamefont {Elcoro}\ \emph {et~al.}(2017)\citenamefont {Elcoro},
  \citenamefont {Bradlyn}, \citenamefont {Wang}, \citenamefont {Vergniory},
  \citenamefont {Cano}, \citenamefont {Felser}, \citenamefont {Bernevig},
  \citenamefont {Orobengoa}, \citenamefont {de~la Flor},\ and\ \citenamefont
  {Aroyo}}]{Elcoro2017}%
  \BibitemOpen
  \bibfield  {author} {\bibinfo {author} {\bibfnamefont {L.}~\bibnamefont
  {Elcoro}}, \bibinfo {author} {\bibfnamefont {B.}~\bibnamefont {Bradlyn}},
  \bibinfo {author} {\bibfnamefont {Z.}~\bibnamefont {Wang}}, \bibinfo {author}
  {\bibfnamefont {M.~G.}\ \bibnamefont {Vergniory}}, \bibinfo {author}
  {\bibfnamefont {J.}~\bibnamefont {Cano}}, \bibinfo {author} {\bibfnamefont
  {C.}~\bibnamefont {Felser}}, \bibinfo {author} {\bibfnamefont {B.~A.}\
  \bibnamefont {Bernevig}}, \bibinfo {author} {\bibfnamefont {D.}~\bibnamefont
  {Orobengoa}}, \bibinfo {author} {\bibfnamefont {G.}~\bibnamefont {de~la
  Flor}},\ and\ \bibinfo {author} {\bibfnamefont {M.~I.}\ \bibnamefont
  {Aroyo}},\ }\bibfield  {title} {\bibinfo {title} {{Double crystallographic
  groups and their representations on the Bilbao Crystallographic Server}},\
  }\href {https://doi.org/10.1107/S1600576717011712} {\bibfield  {journal}
  {\bibinfo  {journal} {Journal of Applied Crystallography}\ }\textbf {\bibinfo
  {volume} {50}},\ \bibinfo {pages} {1457} (\bibinfo {year}
  {2017})}\BibitemShut {NoStop}%
\bibitem [{\citenamefont {Zhang}\ \emph {et~al.}(2019)\citenamefont {Zhang},
  \citenamefont {Jiang}, \citenamefont {Song}, \citenamefont {Huang},
  \citenamefont {He}, \citenamefont {Fang}, \citenamefont {Weng},\ and\
  \citenamefont {Fang}}]{zhang2019catalogue}%
  \BibitemOpen
  \bibfield  {author} {\bibinfo {author} {\bibfnamefont {T.}~\bibnamefont
  {Zhang}}, \bibinfo {author} {\bibfnamefont {Y.}~\bibnamefont {Jiang}},
  \bibinfo {author} {\bibfnamefont {Z.}~\bibnamefont {Song}}, \bibinfo {author}
  {\bibfnamefont {H.}~\bibnamefont {Huang}}, \bibinfo {author} {\bibfnamefont
  {Y.}~\bibnamefont {He}}, \bibinfo {author} {\bibfnamefont {Z.}~\bibnamefont
  {Fang}}, \bibinfo {author} {\bibfnamefont {H.}~\bibnamefont {Weng}},\ and\
  \bibinfo {author} {\bibfnamefont {C.}~\bibnamefont {Fang}},\ }\bibfield
  {title} {\bibinfo {title} {Catalogue of topological electronic materials},\
  }\href@noop {} {\bibfield  {journal} {\bibinfo  {journal} {Nature}\ }\textbf
  {\bibinfo {volume} {566}},\ \bibinfo {pages} {475} (\bibinfo {year}
  {2019})}\BibitemShut {NoStop}%
\bibitem [{\citenamefont {Benalcazar}\ \emph
  {et~al.}(2017{\natexlab{a}})\citenamefont {Benalcazar}, \citenamefont
  {Bernevig},\ and\ \citenamefont {Hughes}}]{Benalcazar2017science}%
  \BibitemOpen
  \bibfield  {author} {\bibinfo {author} {\bibfnamefont {W.~A.}\ \bibnamefont
  {Benalcazar}}, \bibinfo {author} {\bibfnamefont {B.~A.}\ \bibnamefont
  {Bernevig}},\ and\ \bibinfo {author} {\bibfnamefont {T.~L.}\ \bibnamefont
  {Hughes}},\ }\bibfield  {title} {\bibinfo {title} {{Quantized electric
  multipole insulators}},\ }\href {https://doi.org/10.1126/science.aah6442}
  {\bibfield  {journal} {\bibinfo  {journal} {Science}\ }\textbf {\bibinfo
  {volume} {357}},\ \bibinfo {pages} {61} (\bibinfo {year}
  {2017}{\natexlab{a}})},\ \Eprint {https://arxiv.org/abs/1611.07987}
  {arXiv:1611.07987} \BibitemShut {NoStop}%
\bibitem [{\citenamefont {Benalcazar}\ \emph
  {et~al.}(2017{\natexlab{b}})\citenamefont {Benalcazar}, \citenamefont
  {Bernevig},\ and\ \citenamefont {Hughes}}]{Benalcazar2017electric}%
  \BibitemOpen
  \bibfield  {author} {\bibinfo {author} {\bibfnamefont {W.~A.}\ \bibnamefont
  {Benalcazar}}, \bibinfo {author} {\bibfnamefont {B.~A.}\ \bibnamefont
  {Bernevig}},\ and\ \bibinfo {author} {\bibfnamefont {T.~L.}\ \bibnamefont
  {Hughes}},\ }\bibfield  {title} {\bibinfo {title} {{Electric multipole
  moments, topological multipole moment pumping, and chiral hinge states in
  crystalline insulators}},\ }\href
  {https://doi.org/10.1103/PhysRevB.96.245115} {\bibfield  {journal} {\bibinfo
  {journal} {Physical Review B}\ }\textbf {\bibinfo {volume} {96}},\ \bibinfo
  {pages} {245115} (\bibinfo {year} {2017}{\natexlab{b}})}\BibitemShut
  {NoStop}%
\bibitem [{\citenamefont {Khalaf}\ \emph {et~al.}(2018)\citenamefont {Khalaf},
  \citenamefont {Po}, \citenamefont {Vishwanath},\ and\ \citenamefont
  {Watanabe}}]{Khalaf2018symmetry}%
  \BibitemOpen
  \bibfield  {author} {\bibinfo {author} {\bibfnamefont {E.}~\bibnamefont
  {Khalaf}}, \bibinfo {author} {\bibfnamefont {H.~C.}\ \bibnamefont {Po}},
  \bibinfo {author} {\bibfnamefont {A.}~\bibnamefont {Vishwanath}},\ and\
  \bibinfo {author} {\bibfnamefont {H.}~\bibnamefont {Watanabe}},\ }\bibfield
  {title} {\bibinfo {title} {Symmetry indicators and anomalous surface states
  of topological crystalline insulators},\ }\href
  {https://doi.org/10.1103/PhysRevX.8.031070} {\bibfield  {journal} {\bibinfo
  {journal} {Phys. Rev. X}\ }\textbf {\bibinfo {volume} {8}},\ \bibinfo {pages}
  {031070} (\bibinfo {year} {2018})}\BibitemShut {NoStop}%
\bibitem [{\citenamefont {Khalaf}(2018)}]{Khalaf2018higher}%
  \BibitemOpen
  \bibfield  {author} {\bibinfo {author} {\bibfnamefont {E.}~\bibnamefont
  {Khalaf}},\ }\bibfield  {title} {\bibinfo {title} {{Higher-order topological
  insulators and superconductors protected by inversion symmetry}},\ }\href
  {https://doi.org/10.1103/PhysRevB.97.205136} {\bibfield  {journal} {\bibinfo
  {journal} {Physical Review B}\ }\textbf {\bibinfo {volume} {97}},\ \bibinfo
  {pages} {205136} (\bibinfo {year} {2018})},\ \Eprint
  {https://arxiv.org/abs/1801.10050} {arXiv:1801.10050} \BibitemShut {NoStop}%
\bibitem [{\citenamefont {Schindler}\ \emph
  {et~al.}(2018{\natexlab{a}})\citenamefont {Schindler}, \citenamefont {Cook},
  \citenamefont {Vergniory}, \citenamefont {Wang}, \citenamefont {Parkin},
  \citenamefont {Bernevig},\ and\ \citenamefont
  {Neupert}}]{schindler2018higher}%
  \BibitemOpen
  \bibfield  {author} {\bibinfo {author} {\bibfnamefont {F.}~\bibnamefont
  {Schindler}}, \bibinfo {author} {\bibfnamefont {A.~M.}\ \bibnamefont {Cook}},
  \bibinfo {author} {\bibfnamefont {M.~G.}\ \bibnamefont {Vergniory}}, \bibinfo
  {author} {\bibfnamefont {Z.}~\bibnamefont {Wang}}, \bibinfo {author}
  {\bibfnamefont {S.~S.}\ \bibnamefont {Parkin}}, \bibinfo {author}
  {\bibfnamefont {B.~A.}\ \bibnamefont {Bernevig}},\ and\ \bibinfo {author}
  {\bibfnamefont {T.}~\bibnamefont {Neupert}},\ }\bibfield  {title} {\bibinfo
  {title} {Higher-order topological insulators},\ }\href@noop {} {\bibfield
  {journal} {\bibinfo  {journal} {Science advances}\ }\textbf {\bibinfo
  {volume} {4}},\ \bibinfo {pages} {eaat0346} (\bibinfo {year}
  {2018}{\natexlab{a}})}\BibitemShut {NoStop}%
\bibitem [{\citenamefont {Schindler}\ \emph
  {et~al.}(2018{\natexlab{b}})\citenamefont {Schindler}, \citenamefont {Wang},
  \citenamefont {Vergniory}, \citenamefont {Cook}, \citenamefont {Murani},
  \citenamefont {Sengupta}, \citenamefont {Kasumov}, \citenamefont {Deblock},
  \citenamefont {Jeon}, \citenamefont {Drozdov} \emph
  {et~al.}}]{schindler2018higherBi}%
  \BibitemOpen
  \bibfield  {author} {\bibinfo {author} {\bibfnamefont {F.}~\bibnamefont
  {Schindler}}, \bibinfo {author} {\bibfnamefont {Z.}~\bibnamefont {Wang}},
  \bibinfo {author} {\bibfnamefont {M.~G.}\ \bibnamefont {Vergniory}}, \bibinfo
  {author} {\bibfnamefont {A.~M.}\ \bibnamefont {Cook}}, \bibinfo {author}
  {\bibfnamefont {A.}~\bibnamefont {Murani}}, \bibinfo {author} {\bibfnamefont
  {S.}~\bibnamefont {Sengupta}}, \bibinfo {author} {\bibfnamefont {A.~Y.}\
  \bibnamefont {Kasumov}}, \bibinfo {author} {\bibfnamefont {R.}~\bibnamefont
  {Deblock}}, \bibinfo {author} {\bibfnamefont {S.}~\bibnamefont {Jeon}},
  \bibinfo {author} {\bibfnamefont {I.}~\bibnamefont {Drozdov}}, \emph
  {et~al.},\ }\bibfield  {title} {\bibinfo {title} {Higher-order topology in
  bismuth},\ }\href@noop {} {\bibfield  {journal} {\bibinfo  {journal} {Nature
  physics}\ }\textbf {\bibinfo {volume} {14}},\ \bibinfo {pages} {918}
  (\bibinfo {year} {2018}{\natexlab{b}})}\BibitemShut {NoStop}%
\bibitem [{\citenamefont {Song}\ \emph {et~al.}(2017)\citenamefont {Song},
  \citenamefont {Fang},\ and\ \citenamefont {Fang}}]{Song2017}%
  \BibitemOpen
  \bibfield  {author} {\bibinfo {author} {\bibfnamefont {Z.}~\bibnamefont
  {Song}}, \bibinfo {author} {\bibfnamefont {Z.}~\bibnamefont {Fang}},\ and\
  \bibinfo {author} {\bibfnamefont {C.}~\bibnamefont {Fang}},\ }\bibfield
  {title} {\bibinfo {title} {{(d-2)-Dimensional Edge States of Rotation
  Symmetry Protected Topological States}},\ }\href
  {https://doi.org/10.1103/PhysRevLett.119.246402} {\bibfield  {journal}
  {\bibinfo  {journal} {Physical Review Letters}\ }\textbf {\bibinfo {volume}
  {119}},\ \bibinfo {pages} {246402} (\bibinfo {year} {2017})}\BibitemShut
  {NoStop}%
\bibitem [{\citenamefont {Langbehn}\ \emph {et~al.}(2017)\citenamefont
  {Langbehn}, \citenamefont {Peng}, \citenamefont {Trifunovic}, \citenamefont
  {{Von Oppen}},\ and\ \citenamefont {Brouwer}}]{Langbehn2017}%
  \BibitemOpen
  \bibfield  {author} {\bibinfo {author} {\bibfnamefont {J.}~\bibnamefont
  {Langbehn}}, \bibinfo {author} {\bibfnamefont {Y.}~\bibnamefont {Peng}},
  \bibinfo {author} {\bibfnamefont {L.}~\bibnamefont {Trifunovic}}, \bibinfo
  {author} {\bibfnamefont {F.}~\bibnamefont {{Von Oppen}}},\ and\ \bibinfo
  {author} {\bibfnamefont {P.~W.}\ \bibnamefont {Brouwer}},\ }\bibfield
  {title} {\bibinfo {title} {{Reflection-Symmetric Second-Order Topological
  Insulators and Superconductors}},\ }\href
  {https://doi.org/10.1103/PhysRevLett.119.246401} {\bibfield  {journal}
  {\bibinfo  {journal} {Physical Review Letters}\ }\textbf {\bibinfo {volume}
  {119}},\ \bibinfo {pages} {246401} (\bibinfo {year} {2017})},\ \Eprint
  {https://arxiv.org/abs/1708.03640} {arXiv:1708.03640} \BibitemShut {NoStop}%
\bibitem [{\citenamefont {Geier}\ \emph {et~al.}(2018)\citenamefont {Geier},
  \citenamefont {Trifunovic}, \citenamefont {Hoskam},\ and\ \citenamefont
  {Brouwer}}]{Geier2018}%
  \BibitemOpen
  \bibfield  {author} {\bibinfo {author} {\bibfnamefont {M.}~\bibnamefont
  {Geier}}, \bibinfo {author} {\bibfnamefont {L.}~\bibnamefont {Trifunovic}},
  \bibinfo {author} {\bibfnamefont {M.}~\bibnamefont {Hoskam}},\ and\ \bibinfo
  {author} {\bibfnamefont {P.~W.}\ \bibnamefont {Brouwer}},\ }\bibfield
  {title} {\bibinfo {title} {{Second-order topological insulators and
  superconductors with an order-two crystalline symmetry}},\ }\href
  {https://doi.org/10.1103/PhysRevB.97.205135} {\bibfield  {journal} {\bibinfo
  {journal} {Physical Review B}\ }\textbf {\bibinfo {volume} {97}},\ \bibinfo
  {pages} {205135} (\bibinfo {year} {2018})},\ \Eprint
  {https://arxiv.org/abs/1801.10053} {arXiv:1801.10053} \BibitemShut {NoStop}%
\bibitem [{\citenamefont {Imhof}\ \emph {et~al.}(2018)\citenamefont {Imhof},
  \citenamefont {Berger}, \citenamefont {Bayer}, \citenamefont {Brehm},
  \citenamefont {Molenkamp}, \citenamefont {Kiessling}, \citenamefont
  {Schindler}, \citenamefont {Lee}, \citenamefont {Greiter}, \citenamefont
  {Neupert},\ and\ \citenamefont {Thomale}}]{Imhof2017}%
  \BibitemOpen
  \bibfield  {author} {\bibinfo {author} {\bibfnamefont {S.}~\bibnamefont
  {Imhof}}, \bibinfo {author} {\bibfnamefont {C.}~\bibnamefont {Berger}},
  \bibinfo {author} {\bibfnamefont {F.}~\bibnamefont {Bayer}}, \bibinfo
  {author} {\bibfnamefont {J.}~\bibnamefont {Brehm}}, \bibinfo {author}
  {\bibfnamefont {L.~W.}\ \bibnamefont {Molenkamp}}, \bibinfo {author}
  {\bibfnamefont {T.}~\bibnamefont {Kiessling}}, \bibinfo {author}
  {\bibfnamefont {F.}~\bibnamefont {Schindler}}, \bibinfo {author}
  {\bibfnamefont {C.~H.}\ \bibnamefont {Lee}}, \bibinfo {author} {\bibfnamefont
  {M.}~\bibnamefont {Greiter}}, \bibinfo {author} {\bibfnamefont
  {T.}~\bibnamefont {Neupert}},\ and\ \bibinfo {author} {\bibfnamefont
  {R.}~\bibnamefont {Thomale}},\ }\bibfield  {title} {\bibinfo {title}
  {Topolectrical-circuit realization of topological corner modes},\ }\href
  {https://doi.org/10.1038/s41567-018-0246-1} {\bibfield  {journal} {\bibinfo
  {journal} {Nature Physics}\ }\textbf {\bibinfo {volume} {14}},\ \bibinfo
  {pages} {925} (\bibinfo {year} {2018})}\BibitemShut {NoStop}%
\bibitem [{\citenamefont {Peterson}\ \emph {et~al.}(2018)\citenamefont
  {Peterson}, \citenamefont {Benalcazar}, \citenamefont {Hughes},\ and\
  \citenamefont {Bahl}}]{Peterson2018}%
  \BibitemOpen
  \bibfield  {author} {\bibinfo {author} {\bibfnamefont {C.~W.}\ \bibnamefont
  {Peterson}}, \bibinfo {author} {\bibfnamefont {W.~A.}\ \bibnamefont
  {Benalcazar}}, \bibinfo {author} {\bibfnamefont {T.~L.}\ \bibnamefont
  {Hughes}},\ and\ \bibinfo {author} {\bibfnamefont {G.}~\bibnamefont {Bahl}},\
  }\bibfield  {title} {\bibinfo {title} {{A quantized microwave quadrupole
  insulator with topologically protected corner states}},\ }\href
  {https://doi.org/10.1038/nature25777} {\bibfield  {journal} {\bibinfo
  {journal} {Nature}\ }\textbf {\bibinfo {volume} {555}},\ \bibinfo {pages}
  {346} (\bibinfo {year} {2018})}\BibitemShut {NoStop}%
\bibitem [{\citenamefont {Serra-Garcia}\ \emph {et~al.}(2018)\citenamefont
  {Serra-Garcia}, \citenamefont {Peri}, \citenamefont {S{\"{u}}sstrunk},
  \citenamefont {Bilal}, \citenamefont {Larsen}, \citenamefont {Villanueva},\
  and\ \citenamefont {Huber}}]{SerraGarcia2018}%
  \BibitemOpen
  \bibfield  {author} {\bibinfo {author} {\bibfnamefont {M.}~\bibnamefont
  {Serra-Garcia}}, \bibinfo {author} {\bibfnamefont {V.}~\bibnamefont {Peri}},
  \bibinfo {author} {\bibfnamefont {R.}~\bibnamefont {S{\"{u}}sstrunk}},
  \bibinfo {author} {\bibfnamefont {O.~R.}\ \bibnamefont {Bilal}}, \bibinfo
  {author} {\bibfnamefont {T.}~\bibnamefont {Larsen}}, \bibinfo {author}
  {\bibfnamefont {L.~G.}\ \bibnamefont {Villanueva}},\ and\ \bibinfo {author}
  {\bibfnamefont {S.~D.}\ \bibnamefont {Huber}},\ }\bibfield  {title} {\bibinfo
  {title} {{Observation of a phononic quadrupole topological insulator}},\
  }\href {https://doi.org/10.1038/nature25156} {\bibfield  {journal} {\bibinfo
  {journal} {Nature}\ }\textbf {\bibinfo {volume} {555}},\ \bibinfo {pages}
  {342} (\bibinfo {year} {2018})},\ \Eprint {https://arxiv.org/abs/1708.05015}
  {arXiv:1708.05015} \BibitemShut {NoStop}%
\bibitem [{\citenamefont {Noh}\ \emph {et~al.}(2018)\citenamefont {Noh},
  \citenamefont {Benalcazar}, \citenamefont {Huang}, \citenamefont {Collins},
  \citenamefont {Chen}, \citenamefont {Hughes},\ and\ \citenamefont
  {Rechtsman}}]{Noh2018}%
  \BibitemOpen
  \bibfield  {author} {\bibinfo {author} {\bibfnamefont {J.}~\bibnamefont
  {Noh}}, \bibinfo {author} {\bibfnamefont {W.~A.}\ \bibnamefont {Benalcazar}},
  \bibinfo {author} {\bibfnamefont {S.}~\bibnamefont {Huang}}, \bibinfo
  {author} {\bibfnamefont {M.~J.}\ \bibnamefont {Collins}}, \bibinfo {author}
  {\bibfnamefont {K.~P.}\ \bibnamefont {Chen}}, \bibinfo {author}
  {\bibfnamefont {T.~L.}\ \bibnamefont {Hughes}},\ and\ \bibinfo {author}
  {\bibfnamefont {M.~C.}\ \bibnamefont {Rechtsman}},\ }\bibfield  {title}
  {\bibinfo {title} {Topological protection of photonic mid-gap defect modes},\
  }\href {https://doi.org/10.1038/s41566-018-0179-3} {\bibfield  {journal}
  {\bibinfo  {journal} {Nature Photonics}\ }\textbf {\bibinfo {volume} {12}},\
  \bibinfo {pages} {408} (\bibinfo {year} {2018})}\BibitemShut {NoStop}%
\bibitem [{\citenamefont {Trifunovic}\ and\ \citenamefont
  {Brouwer}(2019)}]{Trifunovic2019higher}%
  \BibitemOpen
  \bibfield  {author} {\bibinfo {author} {\bibfnamefont {L.}~\bibnamefont
  {Trifunovic}}\ and\ \bibinfo {author} {\bibfnamefont {P.~W.}\ \bibnamefont
  {Brouwer}},\ }\bibfield  {title} {\bibinfo {title} {Higher-order
  bulk-boundary correspondence for topological crystalline phases},\ }\href
  {https://doi.org/10.1103/PhysRevX.9.011012} {\bibfield  {journal} {\bibinfo
  {journal} {Phys. Rev. X}\ }\textbf {\bibinfo {volume} {9}},\ \bibinfo {pages}
  {011012} (\bibinfo {year} {2019})}\BibitemShut {NoStop}%
\bibitem [{\citenamefont {Queiroz}\ and\ \citenamefont
  {Stern}(2019)}]{queiroz2019splitting}%
  \BibitemOpen
  \bibfield  {author} {\bibinfo {author} {\bibfnamefont {R.}~\bibnamefont
  {Queiroz}}\ and\ \bibinfo {author} {\bibfnamefont {A.}~\bibnamefont
  {Stern}},\ }\bibfield  {title} {\bibinfo {title} {Splitting the hinge mode of
  higher-order topological insulators},\ }\href
  {https://doi.org/10.1103/PhysRevLett.123.036802} {\bibfield  {journal}
  {\bibinfo  {journal} {Phys. Rev. Lett.}\ }\textbf {\bibinfo {volume} {123}},\
  \bibinfo {pages} {036802} (\bibinfo {year} {2019})}\BibitemShut {NoStop}%
\bibitem [{\citenamefont {Fu}\ and\ \citenamefont
  {Kane}(2008)}]{fu2008superconducting}%
  \BibitemOpen
  \bibfield  {author} {\bibinfo {author} {\bibfnamefont {L.}~\bibnamefont
  {Fu}}\ and\ \bibinfo {author} {\bibfnamefont {C.~L.}\ \bibnamefont {Kane}},\
  }\bibfield  {title} {\bibinfo {title} {Superconducting proximity effect and
  majorana fermions at the surface of a topological insulator},\ }\href@noop {}
  {\bibfield  {journal} {\bibinfo  {journal} {Physical review letters}\
  }\textbf {\bibinfo {volume} {100}},\ \bibinfo {pages} {096407} (\bibinfo
  {year} {2008})}\BibitemShut {NoStop}%
\bibitem [{\citenamefont {Nilsson}\ \emph {et~al.}(2008)\citenamefont
  {Nilsson}, \citenamefont {Akhmerov},\ and\ \citenamefont
  {Beenakker}}]{nilsson2008splitting}%
  \BibitemOpen
  \bibfield  {author} {\bibinfo {author} {\bibfnamefont {J.}~\bibnamefont
  {Nilsson}}, \bibinfo {author} {\bibfnamefont {A.}~\bibnamefont {Akhmerov}},\
  and\ \bibinfo {author} {\bibfnamefont {C.}~\bibnamefont {Beenakker}},\
  }\bibfield  {title} {\bibinfo {title} {Splitting of a cooper pair by a pair
  of majorana bound states},\ }\href@noop {} {\bibfield  {journal} {\bibinfo
  {journal} {Physical review letters}\ }\textbf {\bibinfo {volume} {101}},\
  \bibinfo {pages} {120403} (\bibinfo {year} {2008})}\BibitemShut {NoStop}%
\bibitem [{\citenamefont {Fu}\ and\ \citenamefont
  {Kane}(2009)}]{fu2009josephson}%
  \BibitemOpen
  \bibfield  {author} {\bibinfo {author} {\bibfnamefont {L.}~\bibnamefont
  {Fu}}\ and\ \bibinfo {author} {\bibfnamefont {C.~L.}\ \bibnamefont {Kane}},\
  }\bibfield  {title} {\bibinfo {title} {Josephson current and noise at a
  superconductor/quantum-spin-hall-insulator/superconductor junction},\
  }\href@noop {} {\bibfield  {journal} {\bibinfo  {journal} {Physical Review
  B}\ }\textbf {\bibinfo {volume} {79}},\ \bibinfo {pages} {161408} (\bibinfo
  {year} {2009})}\BibitemShut {NoStop}%
\bibitem [{\citenamefont {Tanaka}\ \emph {et~al.}(2009)\citenamefont {Tanaka},
  \citenamefont {Yokoyama},\ and\ \citenamefont
  {Nagaosa}}]{tanaka2009manipulation}%
  \BibitemOpen
  \bibfield  {author} {\bibinfo {author} {\bibfnamefont {Y.}~\bibnamefont
  {Tanaka}}, \bibinfo {author} {\bibfnamefont {T.}~\bibnamefont {Yokoyama}},\
  and\ \bibinfo {author} {\bibfnamefont {N.}~\bibnamefont {Nagaosa}},\
  }\bibfield  {title} {\bibinfo {title} {Manipulation of the majorana fermion,
  andreev reflection, and josephson current on topological insulators},\
  }\href@noop {} {\bibfield  {journal} {\bibinfo  {journal} {Physical review
  letters}\ }\textbf {\bibinfo {volume} {103}},\ \bibinfo {pages} {107002}
  (\bibinfo {year} {2009})}\BibitemShut {NoStop}%
\bibitem [{\citenamefont {Linder}\ \emph {et~al.}(2010)\citenamefont {Linder},
  \citenamefont {Tanaka}, \citenamefont {Yokoyama}, \citenamefont {Sudb{\o}},\
  and\ \citenamefont {Nagaosa}}]{linder2010unconventional}%
  \BibitemOpen
  \bibfield  {author} {\bibinfo {author} {\bibfnamefont {J.}~\bibnamefont
  {Linder}}, \bibinfo {author} {\bibfnamefont {Y.}~\bibnamefont {Tanaka}},
  \bibinfo {author} {\bibfnamefont {T.}~\bibnamefont {Yokoyama}}, \bibinfo
  {author} {\bibfnamefont {A.}~\bibnamefont {Sudb{\o}}},\ and\ \bibinfo
  {author} {\bibfnamefont {N.}~\bibnamefont {Nagaosa}},\ }\bibfield  {title}
  {\bibinfo {title} {Unconventional superconductivity on a topological
  insulator},\ }\href@noop {} {\bibfield  {journal} {\bibinfo  {journal}
  {Physical review letters}\ }\textbf {\bibinfo {volume} {104}},\ \bibinfo
  {pages} {067001} (\bibinfo {year} {2010})}\BibitemShut {NoStop}%
\bibitem [{\citenamefont {Hsu}\ \emph {et~al.}(2018)\citenamefont {Hsu},
  \citenamefont {Stano}, \citenamefont {Klinovaja},\ and\ \citenamefont
  {Loss}}]{PhysRevLett.121.196801}%
  \BibitemOpen
  \bibfield  {author} {\bibinfo {author} {\bibfnamefont {C.-H.}\ \bibnamefont
  {Hsu}}, \bibinfo {author} {\bibfnamefont {P.}~\bibnamefont {Stano}}, \bibinfo
  {author} {\bibfnamefont {J.}~\bibnamefont {Klinovaja}},\ and\ \bibinfo
  {author} {\bibfnamefont {D.}~\bibnamefont {Loss}},\ }\bibfield  {title}
  {\bibinfo {title} {Majorana kramers pairs in higher-order topological
  insulators},\ }\href {https://doi.org/10.1103/PhysRevLett.121.196801}
  {\bibfield  {journal} {\bibinfo  {journal} {Phys. Rev. Lett.}\ }\textbf
  {\bibinfo {volume} {121}},\ \bibinfo {pages} {196801} (\bibinfo {year}
  {2018})}\BibitemShut {NoStop}%
\bibitem [{\citenamefont {Queiroz}\ \emph {et~al.}(2019)\citenamefont
  {Queiroz}, \citenamefont {Fulga}, \citenamefont {Avraham}, \citenamefont
  {Beidenkopf},\ and\ \citenamefont {Cano}}]{queiroz2019partial}%
  \BibitemOpen
  \bibfield  {author} {\bibinfo {author} {\bibfnamefont {R.}~\bibnamefont
  {Queiroz}}, \bibinfo {author} {\bibfnamefont {I.~C.}\ \bibnamefont {Fulga}},
  \bibinfo {author} {\bibfnamefont {N.}~\bibnamefont {Avraham}}, \bibinfo
  {author} {\bibfnamefont {H.}~\bibnamefont {Beidenkopf}},\ and\ \bibinfo
  {author} {\bibfnamefont {J.}~\bibnamefont {Cano}},\ }\bibfield  {title}
  {\bibinfo {title} {Partial lattice defects in higher-order topological
  insulators},\ }\href@noop {} {\bibfield  {journal} {\bibinfo  {journal}
  {Physical Review Letters}\ }\textbf {\bibinfo {volume} {123}},\ \bibinfo
  {pages} {266802} (\bibinfo {year} {2019})}\BibitemShut {NoStop}%
\bibitem [{\citenamefont {Hsieh}\ \emph {et~al.}(2014)\citenamefont {Hsieh},
  \citenamefont {Liu},\ and\ \citenamefont {Fu}}]{hsieh2014topological}%
  \BibitemOpen
  \bibfield  {author} {\bibinfo {author} {\bibfnamefont {T.~H.}\ \bibnamefont
  {Hsieh}}, \bibinfo {author} {\bibfnamefont {J.}~\bibnamefont {Liu}},\ and\
  \bibinfo {author} {\bibfnamefont {L.}~\bibnamefont {Fu}},\ }\bibfield
  {title} {\bibinfo {title} {Topological crystalline insulators and {Dirac}
  octets in antiperovskites},\ }\href@noop {} {\bibfield  {journal} {\bibinfo
  {journal} {Physical Review B}\ }\textbf {\bibinfo {volume} {90}},\ \bibinfo
  {pages} {081112} (\bibinfo {year} {2014})}\BibitemShut {NoStop}%
\bibitem [{\citenamefont {Sun}\ \emph {et~al.}(2010)\citenamefont {Sun},
  \citenamefont {Chen}, \citenamefont {Yunoki}, \citenamefont {Li},\ and\
  \citenamefont {Li}}]{Sun2010}%
  \BibitemOpen
  \bibfield  {author} {\bibinfo {author} {\bibfnamefont {Y.}~\bibnamefont
  {Sun}}, \bibinfo {author} {\bibfnamefont {X.-Q.}\ \bibnamefont {Chen}},
  \bibinfo {author} {\bibfnamefont {S.}~\bibnamefont {Yunoki}}, \bibinfo
  {author} {\bibfnamefont {D.}~\bibnamefont {Li}},\ and\ \bibinfo {author}
  {\bibfnamefont {Y.}~\bibnamefont {Li}},\ }\bibfield  {title} {\bibinfo
  {title} {New family of three-dimensional topological insulators with
  antiperovskite structure},\ }\href
  {https://doi.org/10.1103/PhysRevLett.105.216406} {\bibfield  {journal}
  {\bibinfo  {journal} {Phys. Rev. Lett.}\ }\textbf {\bibinfo {volume} {105}},\
  \bibinfo {pages} {216406} (\bibinfo {year} {2010})}\BibitemShut {NoStop}%
\bibitem [{\citenamefont {Kariyado}\ and\ \citenamefont
  {Ogata}(2011)}]{kariyado2011three}%
  \BibitemOpen
  \bibfield  {author} {\bibinfo {author} {\bibfnamefont {T.}~\bibnamefont
  {Kariyado}}\ and\ \bibinfo {author} {\bibfnamefont {M.}~\bibnamefont
  {Ogata}},\ }\bibfield  {title} {\bibinfo {title} {Three-dimensional dirac
  electrons at the fermi energy in cubic inverse perovskites: Ca3pbo and its
  family},\ }\href@noop {} {\bibfield  {journal} {\bibinfo  {journal} {Journal
  of the Physical Society of Japan}\ }\textbf {\bibinfo {volume} {80}},\
  \bibinfo {pages} {083704} (\bibinfo {year} {2011})}\BibitemShut {NoStop}%
\bibitem [{\citenamefont {Kariyado}\ and\ \citenamefont
  {Ogata}(2012)}]{Kariyado2012}%
  \BibitemOpen
  \bibfield  {author} {\bibinfo {author} {\bibfnamefont {T.}~\bibnamefont
  {Kariyado}}\ and\ \bibinfo {author} {\bibfnamefont {M.}~\bibnamefont
  {Ogata}},\ }\bibfield  {title} {\bibinfo {title} {Low-energy effective
  hamiltonian and the surface states of {Ca$_3$PbO}},\ }\href
  {https://doi.org/10.1143/JPSJ.81.064701} {\bibfield  {journal} {\bibinfo
  {journal} {Journal of the Physical Society of Japan}\ }\textbf {\bibinfo
  {volume} {81}},\ \bibinfo {pages} {064701} (\bibinfo {year} {2012})},\
  \Eprint {https://arxiv.org/abs/https://doi.org/10.1143/JPSJ.81.064701}
  {https://doi.org/10.1143/JPSJ.81.064701} \BibitemShut {NoStop}%
\bibitem [{\citenamefont {Jain}\ \emph {et~al.}(2013)\citenamefont {Jain},
  \citenamefont {Ong}, \citenamefont {Hautier}, \citenamefont {Chen},
  \citenamefont {Richards}, \citenamefont {Dacek}, \citenamefont {Cholia},
  \citenamefont {Gunter}, \citenamefont {Skinner}, \citenamefont {Ceder},\ and\
  \citenamefont {Persson}}]{materialsproject}%
  \BibitemOpen
  \bibfield  {author} {\bibinfo {author} {\bibfnamefont {A.}~\bibnamefont
  {Jain}}, \bibinfo {author} {\bibfnamefont {S.~P.}\ \bibnamefont {Ong}},
  \bibinfo {author} {\bibfnamefont {G.}~\bibnamefont {Hautier}}, \bibinfo
  {author} {\bibfnamefont {W.}~\bibnamefont {Chen}}, \bibinfo {author}
  {\bibfnamefont {W.~D.}\ \bibnamefont {Richards}}, \bibinfo {author}
  {\bibfnamefont {S.}~\bibnamefont {Dacek}}, \bibinfo {author} {\bibfnamefont
  {S.}~\bibnamefont {Cholia}}, \bibinfo {author} {\bibfnamefont
  {D.}~\bibnamefont {Gunter}}, \bibinfo {author} {\bibfnamefont
  {D.}~\bibnamefont {Skinner}}, \bibinfo {author} {\bibfnamefont
  {G.}~\bibnamefont {Ceder}},\ and\ \bibinfo {author} {\bibfnamefont {K.~a.}\
  \bibnamefont {Persson}},\ }\bibfield  {title} {\bibinfo {title} {{The
  Materials Project: A materials genome approach to accelerating materials
  innovation}},\ }\href {https://doi.org/10.1063/1.4812323} {\bibfield
  {journal} {\bibinfo  {journal} {APL Materials}\ }\textbf {\bibinfo {volume}
  {1}},\ \bibinfo {pages} {011002} (\bibinfo {year} {2013})}\BibitemShut
  {NoStop}%
\bibitem [{\citenamefont {Fu}\ \emph {et~al.}(2007)\citenamefont {Fu},
  \citenamefont {Kane},\ and\ \citenamefont {Mele}}]{Fu2007topological}%
  \BibitemOpen
  \bibfield  {author} {\bibinfo {author} {\bibfnamefont {L.}~\bibnamefont
  {Fu}}, \bibinfo {author} {\bibfnamefont {C.~L.}\ \bibnamefont {Kane}},\ and\
  \bibinfo {author} {\bibfnamefont {E.~J.}\ \bibnamefont {Mele}},\ }\bibfield
  {title} {\bibinfo {title} {Topological insulators in three dimensions},\
  }\href {https://doi.org/10.1103/PhysRevLett.98.106803} {\bibfield  {journal}
  {\bibinfo  {journal} {Phys. Rev. Lett.}\ }\textbf {\bibinfo {volume} {98}},\
  \bibinfo {pages} {106803} (\bibinfo {year} {2007})}\BibitemShut {NoStop}%
\bibitem [{\citenamefont {Moore}\ and\ \citenamefont
  {Balents}(2007)}]{Moore2007topological}%
  \BibitemOpen
  \bibfield  {author} {\bibinfo {author} {\bibfnamefont {J.~E.}\ \bibnamefont
  {Moore}}\ and\ \bibinfo {author} {\bibfnamefont {L.}~\bibnamefont
  {Balents}},\ }\bibfield  {title} {\bibinfo {title} {Topological invariants of
  time-reversal-invariant band structures},\ }\href
  {https://doi.org/10.1103/PhysRevB.75.121306} {\bibfield  {journal} {\bibinfo
  {journal} {Phys. Rev. B}\ }\textbf {\bibinfo {volume} {75}},\ \bibinfo
  {pages} {121306} (\bibinfo {year} {2007})}\BibitemShut {NoStop}%
\bibitem [{\citenamefont {Roy}(2009)}]{Roy2009topological}%
  \BibitemOpen
  \bibfield  {author} {\bibinfo {author} {\bibfnamefont {R.}~\bibnamefont
  {Roy}},\ }\bibfield  {title} {\bibinfo {title} {Topological phases and the
  quantum spin hall effect in three dimensions},\ }\href
  {https://doi.org/10.1103/PhysRevB.79.195322} {\bibfield  {journal} {\bibinfo
  {journal} {Phys. Rev. B}\ }\textbf {\bibinfo {volume} {79}},\ \bibinfo
  {pages} {195322} (\bibinfo {year} {2009})}\BibitemShut {NoStop}%
\bibitem [{\citenamefont {Fu}\ and\ \citenamefont {Kane}(2007)}]{Fu_2007}%
  \BibitemOpen
  \bibfield  {author} {\bibinfo {author} {\bibfnamefont {L.}~\bibnamefont
  {Fu}}\ and\ \bibinfo {author} {\bibfnamefont {C.~L.}\ \bibnamefont {Kane}},\
  }\bibfield  {title} {\bibinfo {title} {Topological insulators with inversion
  symmetry},\ }\bibfield  {journal} {\bibinfo  {journal} {Physical Review B}\
  }\textbf {\bibinfo {volume} {76}},\ \href
  {https://doi.org/10.1103/physrevb.76.045302} {10.1103/physrevb.76.045302}
  (\bibinfo {year} {2007})\BibitemShut {NoStop}%
\bibitem [{\citenamefont {Chiu}\ \emph {et~al.}(2017)\citenamefont {Chiu},
  \citenamefont {Chan}, \citenamefont {Li}, \citenamefont {Nohara},\ and\
  \citenamefont {Schnyder}}]{Chiu2017typeII}%
  \BibitemOpen
  \bibfield  {author} {\bibinfo {author} {\bibfnamefont {C.-K.}\ \bibnamefont
  {Chiu}}, \bibinfo {author} {\bibfnamefont {Y.-H.}\ \bibnamefont {Chan}},
  \bibinfo {author} {\bibfnamefont {X.}~\bibnamefont {Li}}, \bibinfo {author}
  {\bibfnamefont {Y.}~\bibnamefont {Nohara}},\ and\ \bibinfo {author}
  {\bibfnamefont {A.~P.}\ \bibnamefont {Schnyder}},\ }\bibfield  {title}
  {\bibinfo {title} {Type-ii dirac surface states in topological crystalline
  insulators},\ }\href {https://doi.org/10.1103/PhysRevB.95.035151} {\bibfield
  {journal} {\bibinfo  {journal} {Phys. Rev. B}\ }\textbf {\bibinfo {volume}
  {95}},\ \bibinfo {pages} {035151} (\bibinfo {year} {2017})}\BibitemShut
  {NoStop}%
\bibitem [{\citenamefont {Teo}\ \emph {et~al.}(2008)\citenamefont {Teo},
  \citenamefont {Fu},\ and\ \citenamefont {Kane}}]{Teo2008surface}%
  \BibitemOpen
  \bibfield  {author} {\bibinfo {author} {\bibfnamefont {J.~C.~Y.}\
  \bibnamefont {Teo}}, \bibinfo {author} {\bibfnamefont {L.}~\bibnamefont
  {Fu}},\ and\ \bibinfo {author} {\bibfnamefont {C.~L.}\ \bibnamefont {Kane}},\
  }\bibfield  {title} {\bibinfo {title} {Surface states and topological
  invariants in three-dimensional topological insulators: Application to
  ${\text{bi}}_{1\ensuremath{-}x}{\text{sb}}_{x}$},\ }\href
  {https://doi.org/10.1103/PhysRevB.78.045426} {\bibfield  {journal} {\bibinfo
  {journal} {Phys. Rev. B}\ }\textbf {\bibinfo {volume} {78}},\ \bibinfo
  {pages} {045426} (\bibinfo {year} {2008})}\BibitemShut {NoStop}%
\bibitem [{pyt()}]{pythtb}%
  \BibitemOpen
  \href@noop {} {\bibinfo {title} {{S. Coh, D. Vanderbilt} python tight binding
  (pythtb) (2013)}},\ \bibinfo {howpublished}
  {\url{www.physics.rutgers.edu/pythtb}}\BibitemShut {NoStop}%
\bibitem [{\citenamefont {Kawakami}\ \emph {et~al.}(2018)\citenamefont
  {Kawakami}, \citenamefont {Okamura}, \citenamefont {Kobayashi},\ and\
  \citenamefont {Sato}}]{PhysRevX.8.041026}%
  \BibitemOpen
  \bibfield  {author} {\bibinfo {author} {\bibfnamefont {T.}~\bibnamefont
  {Kawakami}}, \bibinfo {author} {\bibfnamefont {T.}~\bibnamefont {Okamura}},
  \bibinfo {author} {\bibfnamefont {S.}~\bibnamefont {Kobayashi}},\ and\
  \bibinfo {author} {\bibfnamefont {M.}~\bibnamefont {Sato}},\ }\bibfield
  {title} {\bibinfo {title} {Topological crystalline materials of $j=3/2$
  electrons: Antiperovskites, dirac points, and high winding topological
  superconductivity},\ }\href {https://doi.org/10.1103/PhysRevX.8.041026}
  {\bibfield  {journal} {\bibinfo  {journal} {Phys. Rev. X}\ }\textbf {\bibinfo
  {volume} {8}},\ \bibinfo {pages} {041026} (\bibinfo {year}
  {2018})}\BibitemShut {NoStop}%
\bibitem [{\citenamefont {Pertsova}\ \emph {et~al.}(2019)\citenamefont
  {Pertsova}, \citenamefont {Geilhufe}, \citenamefont {Bremholm},\ and\
  \citenamefont {Balatsky}}]{PhysRevB.99.205126}%
  \BibitemOpen
  \bibfield  {author} {\bibinfo {author} {\bibfnamefont {A.}~\bibnamefont
  {Pertsova}}, \bibinfo {author} {\bibfnamefont {R.~M.}\ \bibnamefont
  {Geilhufe}}, \bibinfo {author} {\bibfnamefont {M.}~\bibnamefont {Bremholm}},\
  and\ \bibinfo {author} {\bibfnamefont {A.~V.}\ \bibnamefont {Balatsky}},\
  }\bibfield  {title} {\bibinfo {title} {Computational search for dirac and
  weyl nodes in $f$-electron antiperovskites},\ }\href
  {https://doi.org/10.1103/PhysRevB.99.205126} {\bibfield  {journal} {\bibinfo
  {journal} {Phys. Rev. B}\ }\textbf {\bibinfo {volume} {99}},\ \bibinfo
  {pages} {205126} (\bibinfo {year} {2019})}\BibitemShut {NoStop}%
\bibitem [{\citenamefont {Oudah}\ \emph {et~al.}(2016)\citenamefont {Oudah},
  \citenamefont {Ikeda}, \citenamefont {Hausmann}, \citenamefont {Yonezawa},
  \citenamefont {Fukumoto}, \citenamefont {Kobayashi}, \citenamefont {Sato},\
  and\ \citenamefont {Maeno}}]{Oudah_2016}%
  \BibitemOpen
  \bibfield  {author} {\bibinfo {author} {\bibfnamefont {M.}~\bibnamefont
  {Oudah}}, \bibinfo {author} {\bibfnamefont {A.}~\bibnamefont {Ikeda}},
  \bibinfo {author} {\bibfnamefont {J.~N.}\ \bibnamefont {Hausmann}}, \bibinfo
  {author} {\bibfnamefont {S.}~\bibnamefont {Yonezawa}}, \bibinfo {author}
  {\bibfnamefont {T.}~\bibnamefont {Fukumoto}}, \bibinfo {author}
  {\bibfnamefont {S.}~\bibnamefont {Kobayashi}}, \bibinfo {author}
  {\bibfnamefont {M.}~\bibnamefont {Sato}},\ and\ \bibinfo {author}
  {\bibfnamefont {Y.}~\bibnamefont {Maeno}},\ }\bibfield  {title} {\bibinfo
  {title} {Superconductivity in the antiperovskite {Dirac}-metal oxide
  {Sr$_{3−x}$SnO}},\ }\bibfield  {journal} {\bibinfo  {journal} {Nature
  Communications}\ }\textbf {\bibinfo {volume} {7}},\ \href
  {https://doi.org/10.1038/ncomms13617} {10.1038/ncomms13617} (\bibinfo {year}
  {2016})\BibitemShut {NoStop}%
\bibitem [{\citenamefont {Ma}\ \emph {et~al.}(2019)\citenamefont {Ma},
  \citenamefont {Edgeton}, \citenamefont {Paik}, \citenamefont {Faeth},
  \citenamefont {Parzyck}, \citenamefont {Pamuk}, \citenamefont {Shang},
  \citenamefont {Liu}, \citenamefont {Shen}, \citenamefont {Schlom} \emph
  {et~al.}}]{ma2019realization}%
  \BibitemOpen
  \bibfield  {author} {\bibinfo {author} {\bibfnamefont {Y.}~\bibnamefont
  {Ma}}, \bibinfo {author} {\bibfnamefont {A.}~\bibnamefont {Edgeton}},
  \bibinfo {author} {\bibfnamefont {H.}~\bibnamefont {Paik}}, \bibinfo {author}
  {\bibfnamefont {B.}~\bibnamefont {Faeth}}, \bibinfo {author} {\bibfnamefont
  {C.}~\bibnamefont {Parzyck}}, \bibinfo {author} {\bibfnamefont
  {B.}~\bibnamefont {Pamuk}}, \bibinfo {author} {\bibfnamefont {S.-L.}\
  \bibnamefont {Shang}}, \bibinfo {author} {\bibfnamefont {Z.-K.}\ \bibnamefont
  {Liu}}, \bibinfo {author} {\bibfnamefont {K.~M.}\ \bibnamefont {Shen}},
  \bibinfo {author} {\bibfnamefont {D.~G.}\ \bibnamefont {Schlom}}, \emph
  {et~al.},\ }\bibfield  {title} {\bibinfo {title} {Realization of epitaxial
  thin films of the topological crystalline insulator {Sr$_3$SnO}},\
  }\href@noop {} {\bibfield  {journal} {\bibinfo  {journal} {arXiv preprint
  arXiv:1912.13431}\ } (\bibinfo {year} {2019})}\BibitemShut {NoStop}%
\bibitem [{\citenamefont {Suetsugu}\ \emph {et~al.}(2018)\citenamefont
  {Suetsugu}, \citenamefont {Hayama}, \citenamefont {Rost}, \citenamefont
  {Nuss}, \citenamefont {Mühle}, \citenamefont {Kim}, \citenamefont
  {Kitagawa},\ and\ \citenamefont {Takagi}}]{Suetsugu_2018}%
  \BibitemOpen
  \bibfield  {author} {\bibinfo {author} {\bibfnamefont {S.}~\bibnamefont
  {Suetsugu}}, \bibinfo {author} {\bibfnamefont {K.}~\bibnamefont {Hayama}},
  \bibinfo {author} {\bibfnamefont {A.~W.}\ \bibnamefont {Rost}}, \bibinfo
  {author} {\bibfnamefont {J.}~\bibnamefont {Nuss}}, \bibinfo {author}
  {\bibfnamefont {C.}~\bibnamefont {Mühle}}, \bibinfo {author} {\bibfnamefont
  {J.}~\bibnamefont {Kim}}, \bibinfo {author} {\bibfnamefont {K.}~\bibnamefont
  {Kitagawa}},\ and\ \bibinfo {author} {\bibfnamefont {H.}~\bibnamefont
  {Takagi}},\ }\bibfield  {title} {\bibinfo {title} {Magnetotransport in
  {Sr$_3$PbO} antiperovskite},\ }\bibfield  {journal} {\bibinfo  {journal}
  {Physical Review B}\ }\textbf {\bibinfo {volume} {98}},\ \href
  {https://doi.org/10.1103/physrevb.98.115203} {10.1103/physrevb.98.115203}
  (\bibinfo {year} {2018})\BibitemShut {NoStop}%
\bibitem [{\citenamefont {Obata}\ \emph {et~al.}(2017)\citenamefont {Obata},
  \citenamefont {Yukawa}, \citenamefont {Horiba}, \citenamefont {Kumigashira},
  \citenamefont {Toda}, \citenamefont {Matsuishi},\ and\ \citenamefont
  {Hosono}}]{PhysRevB.96.155109}%
  \BibitemOpen
  \bibfield  {author} {\bibinfo {author} {\bibfnamefont {Y.}~\bibnamefont
  {Obata}}, \bibinfo {author} {\bibfnamefont {R.}~\bibnamefont {Yukawa}},
  \bibinfo {author} {\bibfnamefont {K.}~\bibnamefont {Horiba}}, \bibinfo
  {author} {\bibfnamefont {H.}~\bibnamefont {Kumigashira}}, \bibinfo {author}
  {\bibfnamefont {Y.}~\bibnamefont {Toda}}, \bibinfo {author} {\bibfnamefont
  {S.}~\bibnamefont {Matsuishi}},\ and\ \bibinfo {author} {\bibfnamefont
  {H.}~\bibnamefont {Hosono}},\ }\bibfield  {title} {\bibinfo {title} {Arpes
  studies of the inverse perovskite {Ca$_3$PbO}: Experimental confirmation of a
  candidate 3d dirac fermion system},\ }\href
  {https://doi.org/10.1103/PhysRevB.96.155109} {\bibfield  {journal} {\bibinfo
  {journal} {Phys. Rev. B}\ }\textbf {\bibinfo {volume} {96}},\ \bibinfo
  {pages} {155109} (\bibinfo {year} {2017})}\BibitemShut {NoStop}%
\bibitem [{\citenamefont {Kitagawa}\ \emph {et~al.}(2018)\citenamefont
  {Kitagawa}, \citenamefont {Ishida}, \citenamefont {Oudah}, \citenamefont
  {Hausmann}, \citenamefont {Ikeda}, \citenamefont {Yonezawa},\ and\
  \citenamefont {Maeno}}]{Kitagawa_2018}%
  \BibitemOpen
  \bibfield  {author} {\bibinfo {author} {\bibfnamefont {S.}~\bibnamefont
  {Kitagawa}}, \bibinfo {author} {\bibfnamefont {K.}~\bibnamefont {Ishida}},
  \bibinfo {author} {\bibfnamefont {M.}~\bibnamefont {Oudah}}, \bibinfo
  {author} {\bibfnamefont {J.~N.}\ \bibnamefont {Hausmann}}, \bibinfo {author}
  {\bibfnamefont {A.}~\bibnamefont {Ikeda}}, \bibinfo {author} {\bibfnamefont
  {S.}~\bibnamefont {Yonezawa}},\ and\ \bibinfo {author} {\bibfnamefont
  {Y.}~\bibnamefont {Maeno}},\ }\bibfield  {title} {\bibinfo {title}
  {Normal-state properties of the antiperovskite oxide {Sr$_{3−x}$SnO}
  revealed by {Sn119} -nmr},\ }\bibfield  {journal} {\bibinfo  {journal}
  {Physical Review B}\ }\textbf {\bibinfo {volume} {98}},\ \href
  {https://doi.org/10.1103/physrevb.98.100503} {10.1103/physrevb.98.100503}
  (\bibinfo {year} {2018})\BibitemShut {NoStop}%
\bibitem [{\citenamefont {Rost}\ \emph {et~al.}(2019)\citenamefont {Rost},
  \citenamefont {Kim}, \citenamefont {Suetsugu}, \citenamefont {Abdolazimi},
  \citenamefont {Hayama}, \citenamefont {Bruin}, \citenamefont {Mühle},
  \citenamefont {Kitagawa}, \citenamefont {Yaresko}, \citenamefont {Nuss},\
  and\ \citenamefont {Takagi}}]{Rost2019inverse}%
  \BibitemOpen
  \bibfield  {author} {\bibinfo {author} {\bibfnamefont {A.~W.}\ \bibnamefont
  {Rost}}, \bibinfo {author} {\bibfnamefont {J.}~\bibnamefont {Kim}}, \bibinfo
  {author} {\bibfnamefont {S.}~\bibnamefont {Suetsugu}}, \bibinfo {author}
  {\bibfnamefont {V.}~\bibnamefont {Abdolazimi}}, \bibinfo {author}
  {\bibfnamefont {K.}~\bibnamefont {Hayama}}, \bibinfo {author} {\bibfnamefont
  {J.~A.~N.}\ \bibnamefont {Bruin}}, \bibinfo {author} {\bibfnamefont
  {C.}~\bibnamefont {Mühle}}, \bibinfo {author} {\bibfnamefont
  {K.}~\bibnamefont {Kitagawa}}, \bibinfo {author} {\bibfnamefont
  {A.}~\bibnamefont {Yaresko}}, \bibinfo {author} {\bibfnamefont
  {J.}~\bibnamefont {Nuss}},\ and\ \bibinfo {author} {\bibfnamefont
  {H.}~\bibnamefont {Takagi}},\ }\bibfield  {title} {\bibinfo {title}
  {Inverse-perovskites a3bo (a = sr, ca, eu/b = pb, sn): A platform for control
  of dirac and weyl fermions},\ }\href {https://doi.org/10.1063/1.5129695}
  {\bibfield  {journal} {\bibinfo  {journal} {APL Materials}\ }\textbf
  {\bibinfo {volume} {7}},\ \bibinfo {pages} {121114} (\bibinfo {year}
  {2019})},\ \Eprint {https://arxiv.org/abs/https://doi.org/10.1063/1.5129695}
  {https://doi.org/10.1063/1.5129695} \BibitemShut {NoStop}%
\bibitem [{\citenamefont {Huang}\ \emph {et~al.}(2019)\citenamefont {Huang},
  \citenamefont {Nakamura}, \citenamefont {K\"uster}, \citenamefont {Yaresko},
  \citenamefont {Samal}, \citenamefont {Schr\"oter}, \citenamefont {Strocov},
  \citenamefont {Starke},\ and\ \citenamefont {Takagi}}]{Huang2019unusual}%
  \BibitemOpen
  \bibfield  {author} {\bibinfo {author} {\bibfnamefont {D.}~\bibnamefont
  {Huang}}, \bibinfo {author} {\bibfnamefont {H.}~\bibnamefont {Nakamura}},
  \bibinfo {author} {\bibfnamefont {K.}~\bibnamefont {K\"uster}}, \bibinfo
  {author} {\bibfnamefont {A.}~\bibnamefont {Yaresko}}, \bibinfo {author}
  {\bibfnamefont {D.}~\bibnamefont {Samal}}, \bibinfo {author} {\bibfnamefont
  {N.~B.~M.}\ \bibnamefont {Schr\"oter}}, \bibinfo {author} {\bibfnamefont
  {V.~N.}\ \bibnamefont {Strocov}}, \bibinfo {author} {\bibfnamefont
  {U.}~\bibnamefont {Starke}},\ and\ \bibinfo {author} {\bibfnamefont
  {H.}~\bibnamefont {Takagi}},\ }\bibfield  {title} {\bibinfo {title} {Unusual
  valence state in the antiperovskites ${\mathrm{sr}}_{3}\mathrm{SnO}$ and
  ${\mathrm{sr}}_{3}\mathrm{PbO}$ revealed by x-ray photoelectron
  spectroscopy},\ }\href {https://doi.org/10.1103/PhysRevMaterials.3.124203}
  {\bibfield  {journal} {\bibinfo  {journal} {Phys. Rev. Materials}\ }\textbf
  {\bibinfo {volume} {3}},\ \bibinfo {pages} {124203} (\bibinfo {year}
  {2019})}\BibitemShut {NoStop}%
\bibitem [{\citenamefont {Jackiw}\ and\ \citenamefont
  {Rebbi}(1976)}]{jackiw1976solitons}%
  \BibitemOpen
  \bibfield  {author} {\bibinfo {author} {\bibfnamefont {R.}~\bibnamefont
  {Jackiw}}\ and\ \bibinfo {author} {\bibfnamefont {C.}~\bibnamefont {Rebbi}},\
  }\bibfield  {title} {\bibinfo {title} {Solitons with fermion number $1/2$},\
  }\href@noop {} {\bibfield  {journal} {\bibinfo  {journal} {Physical Review
  D}\ }\textbf {\bibinfo {volume} {13}},\ \bibinfo {pages} {3398} (\bibinfo
  {year} {1976})}\BibitemShut {NoStop}%
\bibitem [{\citenamefont {Hsieh}\ \emph {et~al.}(2012)\citenamefont {Hsieh},
  \citenamefont {Lin}, \citenamefont {Liu}, \citenamefont {Duan}, \citenamefont
  {Bansil},\ and\ \citenamefont {Fu}}]{hsieh2012topological}%
  \BibitemOpen
  \bibfield  {author} {\bibinfo {author} {\bibfnamefont {T.~H.}\ \bibnamefont
  {Hsieh}}, \bibinfo {author} {\bibfnamefont {H.}~\bibnamefont {Lin}}, \bibinfo
  {author} {\bibfnamefont {J.}~\bibnamefont {Liu}}, \bibinfo {author}
  {\bibfnamefont {W.}~\bibnamefont {Duan}}, \bibinfo {author} {\bibfnamefont
  {A.}~\bibnamefont {Bansil}},\ and\ \bibinfo {author} {\bibfnamefont
  {L.}~\bibnamefont {Fu}},\ }\bibfield  {title} {\bibinfo {title} {Topological
  crystalline insulators in the {SnTe} material class},\ }\href@noop {}
  {\bibfield  {journal} {\bibinfo  {journal} {Nature communications}\ }\textbf
  {\bibinfo {volume} {3}},\ \bibinfo {pages} {982} (\bibinfo {year}
  {2012})}\BibitemShut {NoStop}%
\end{thebibliography}%


%

\appendix

\section{Spin 3/2 matrices\label{jmatrix}}
Here we define the spin 3/2 matrices $\bm{J}$ and $\bm{\tilde{J}}$ that were used in our Hamiltonian (\ref{Hameqn}). \par
\begin{align}
    J_{x} &= \left(\begin{array}{cccc}{0} & {\frac{\sqrt{3}}{2}} & {0} & {0} \\ {\frac{\sqrt{3}}{2}} & {0} & {1} & {0} \\ 
    {0} & {1} & {0} & {\frac{\sqrt{3}}{2}} \\ 
    {0} & {0} & {\frac{\sqrt{3}}{2}} & {0}\end{array}\right)
\\
    J_{y} &= \left(\begin{array}{cccc}{0} & {-i \frac{\sqrt{3}}{2}} & {0} & {0} \\ {i \frac{\sqrt{3}}{2}} & {0} & {-i} & {0} \\ {0} & {i} & {0} & {-i \frac{\sqrt{3}}{2}} \\ {0} & {0} & {i \frac{\sqrt{3}}{2}} & {0}\end{array}\right)
\\
    J_{z} &= \left(\begin{array}{cccc}{\frac{3}{2}} & {0} & {0} & {0} \\ {0} & {\frac{1}{2}} & {0} & {0} \\ {0} & {0} & {-\frac{1}{2}} & {0} \\ {0} & {0} & {0} & {-\frac{3}{2}}\end{array}\right)
\end{align}

\begin{align}
    \tilde{J}_{x} &= \left(\begin{array}{cccc}{0} & {\frac{\sqrt{3}}{4}} & {0} & {-\frac{5}{4}} \\ {\frac{\sqrt{3}}{4}} & {0} & {-\frac{3}{4}} & {0} \\ {0} & {-\frac{3}{4}} & {0} & {\frac{\sqrt{3}}{4}} \\ {-\frac{5}{4}} & {0} & {\frac{\sqrt{3}}{4}} & {0}\end{array}\right)
\\
    \tilde{J}_{y} &= \left(\begin{array}{cccc}{0} & {-i \frac{\sqrt{3}}{4}} & {0} & {-i \frac{5}{4}} \\ {i \frac{\sqrt{3}}{4}} & {0} & {i \frac{3}{4}} & {0} \\ {0} & {-i \frac{3}{4}} & {0} & {-i \frac{\sqrt{3}}{4}} \\ {i \frac{5}{4}} & {0} & {i \frac{\sqrt{3}}{4}} & {0}\end{array}\right)
\\
    \tilde{J}_{z} &= \left(\begin{array}{cccc}{-\frac{1}{2}} & {0} & {0} & {0} \\ {0} & {\frac{3}{2}} & {0} & {0} \\ {0} & {0} & {-\frac{3}{2}} & {0} \\ {0} & {0} & {0} & {\frac{1}{2}}\end{array}\right)
\end{align}

\section{Surface theory\label{surface_theory}}
The $k\cdot p$ Hamiltonian that describes the low-energy physics of phase (I) near the $\Gamma$ point can be derived by expanding Eq.~(\ref{Hameqn}) in powers of $k$.
We divide the Hamiltonian into a linear order part and a quadratic part:
\begin{align}
    &H_{{k.p}} =~ H_{linear} + H_{quadratic}\\
    &H_{linear} =~ m\sigma_0\tau_0\rho_z+v_1 \bm{k}\cdot \bm{J}+v_2 \bm{k}\cdot \bm{\tilde{J}}\\
    &H_{quadratic} =
    \begin{bmatrix}
   H_{{0k}}(\alpha_{1},\beta_{1},\gamma_{1}) &
   0 \\
   0&
   H_{{0k}}(
  \alpha_{2},\beta_{2},\gamma_{2}).
   \end{bmatrix} \label{eqn_quadratic}\\
   &H_{0k}(\alpha,\beta,\gamma) = \frac{\alpha}{4}k^2\sigma_0\tau_0+\frac{\beta}{8}(2k_z^2-k_x^2-k_y^2)\sigma_z\tau_z+\nonumber \\
   &\frac{\sqrt{3}\beta}{8}(k_x^2-k_y^2)\sigma_0\tau_x  +\gamma (k_x k_y \sigma_x\tau_z+k_y k_z \sigma_y\tau_z+k_x k_y \sigma_0\tau_y),
\end{align}
where $\bm{J}$ and $\bm{\tilde{J}}$ are explicitly shown in Appendix \ref{jmatrix}. We presume that the quadratic Hamiltonian is much smaller than the linear Hamiltonian, so that we can treat it as a perturbation term. Then the mass term $m$ controls the band inversion at $\Gamma$, i.e., $m>0$ indicates a trivial state and $m<0$ indicates a double-band inversion at $\Gamma$.
We analytically derive the low-energy surface theory following Jackiw-Rebbi \cite{jackiw1976solitons, PhysRevB.97.205135,PhysRevX.9.011012}.
There are three useful variables defined by Eq.~(\ref{eqn_v}) and (\ref{eqn_R}): $v_d = (v_1-2v_2)/2$, $v_s = (2v_1+v_2)/2$, and $R = v_d/v_s$. \par

We will utilize the bulk-boundary correspondence to calculate the mirror Chern number by counting the surface states that cross the Fermi level: 
\par
\begin{equation}
\label{eqncn}
    C_m = \frac{N_R^+-N_L^+-N_R^-+N_L^-}{2}=N_R^+-N_L^+,
\end{equation}
where $N_{R/L}^\pm$ indicates the number of surface states that are right (R)- or left (L)-moving (specifically, positive or negative slope) in the sector with $\pm i$ mirror eigenvalues.
The second equality follows from time reversal symmetry.

\subsection{$M_z$ mirror Chern surface states}
\subsubsection{Mirror Chern number}
In momentum space, $k_z=0$ is a $M_z$ invariant plane, where the Hamiltonian can be decomposed into two mirror-invariant subspaces. The mirror operator $M_z = C_{2z} {\cal P}$ has eigenvalues $\pm i$.  The eigenvectors provide a unitary basis transformation matrix,
\begin{equation}
    U = \left(
    \begin{smallmatrix}
    0 & 0 & 0 & 0 & 0 & 0 & 0 & 1 \\
 0 & 0 & 0 & 1 & 0 & 0 & 0 & 0 \\
 0 & 0 & 0 & 0 & 0 & 0 & 1 & 0 \\
 0 & 0 & 1 & 0 & 0 & 0 & 0 & 0 \\
 0 & 1 & 0 & 0 & 0 & 0 & 0 & 0 \\
 0 & 0 & 0 & 0 & 0 & 1 & 0 & 0 \\
 1 & 0 & 0 & 0 & 0 & 0 & 0 & 0 \\
    \end{smallmatrix}    
    \right)
\end{equation}
After conjugating by $U$, the linear order Hamiltonian $H_{linear}$ is expressed as $H_{M_z}^{+i}\oplus H_{M_z}^{-i}$, where the superscript indicates the mirror eigenvalue of each block. $H_{M_z}^{\pm i}$ are each functions of $k_x$ and $k_y$, with $k_z=0$. The expression for $H_{M_z}^{\pm i}$ is
\begin{align}
\label{ham_mz}
    H_{M_z}^{\pm i} = -m\tau_z \mp \left( \frac{\sqrt{3}}{2}v_s\tau_y-v_d\sigma_x\tau_y \pm \frac{v_s}{2}\sigma_y\tau_x \right)k_y  \nonumber \\
 +\left(\frac{\sqrt{3}}{2}v_s\tau_x+v_d\sigma_x\tau_x \pm \frac{v_s}{2}\sigma_y\tau_y\right)k_x
\end{align}

Now consider a slab that is finite in the $y$ direction but infinite in the $x$ and $z$ directions. Then we can express the mass $m(y)$ as a function of $y$, which changes sign across $y=0$.
To solve for the eigenstates of the slab, we replace $k_y$ with $-i\partial_y$ to get a continuum surface model for the $k_z = 0$ mirror plane. 
Zero energy eigenstates will be solutions to the Schr\"{o}dinger equation $H_{M_z}^{\pm i}\Psi =0 $. Although after the substitution $k_y \rightarrow -i\partial_y$, $H_{M_z}^{\pm i}$ is not Hermitian any more, we can nonetheless find real eigenvalues that correspond to localized boundary states.

To do this, we use the Jackiw-Rebbi ansatz \cite{jackiw1976solitons} and start with a solution of the form: $\Psi = {\cal  N}e^{-\frac{1}{\xi}\int_0^{y}m(y)dy} \chi$. Here $\cal  N$ is a normalization factor, $\xi$ is the exponential decay length, and $\chi$ is a spinor. 
Plugging $\Psi$ into the Schr\"{o}dinger equation with the Hamiltonian $H_{M_z}^{\pm i}$ and taking $k_x=0$ yields the equation:
\begin{equation}
    \{-m\tau_z \mp \left( \frac{\sqrt{3}}{2}v_s\tau_y-v_d\tau_y\pm\frac{v_s}{2}\sigma_y\tau_x \right)(-i\frac{-m}{\xi})\}\chi = 0
\end{equation}
Let us first focus on the $+i$ subspace. There are four solutions that satisfy the equation, with four different decay lengths $\xi = \pm(v_d\pm v_s)$. However, the solution should be normalizable, which requires $\xi > 0$. If we choose $v_d> v_s>0$ then the two physical solutions are: $\xi_1 = v_d+v_s$, $\chi_1 = \frac{1}{2\sqrt{2}}(\sqrt{3},-1,-1,\sqrt{3})$, and $\xi_2 = v_d-v_s$, $\chi_2 = \frac{1}{2\sqrt{2}}(1,\sqrt{3},\sqrt{3},1)$. \par
We then project the remaining second line of our Hamiltonian (\ref{ham_mz}) into the low energy subspace spanned by the two solutions, which yields: 
\begin{equation}
    \langle \chi_i|H_{M_z}^{+i}|\chi_j \rangle = 
    \begin{pmatrix}
    -(v_d-\frac{v_s}{2})k_x & \frac{\sqrt{3}}{2}v_s k _x \\   \frac{\sqrt{3}}{2}v_s k _x & -(v_d+\frac{v_s}{2})k_x
 \end{pmatrix}
\end{equation}
The energy eigenvalues are all linear in $k_x$ and have negative slopes $-(v_d\pm v_s)k_x$. Similarly, for the $-i$ subspace, the energy eigenvalues are $(v_d\pm v_s)k_x$. This result can be understood because the $-i$ bands are the Kramers partners of $+i$ bands. For the case $v_d<v_s<0$, the same result can be obtained. Thus, we deduce that for the case $R>1$ the mirror Chern number is $C_m=-2$.\par

\begin{figure*}[t]
\includegraphics[width = 16cm]{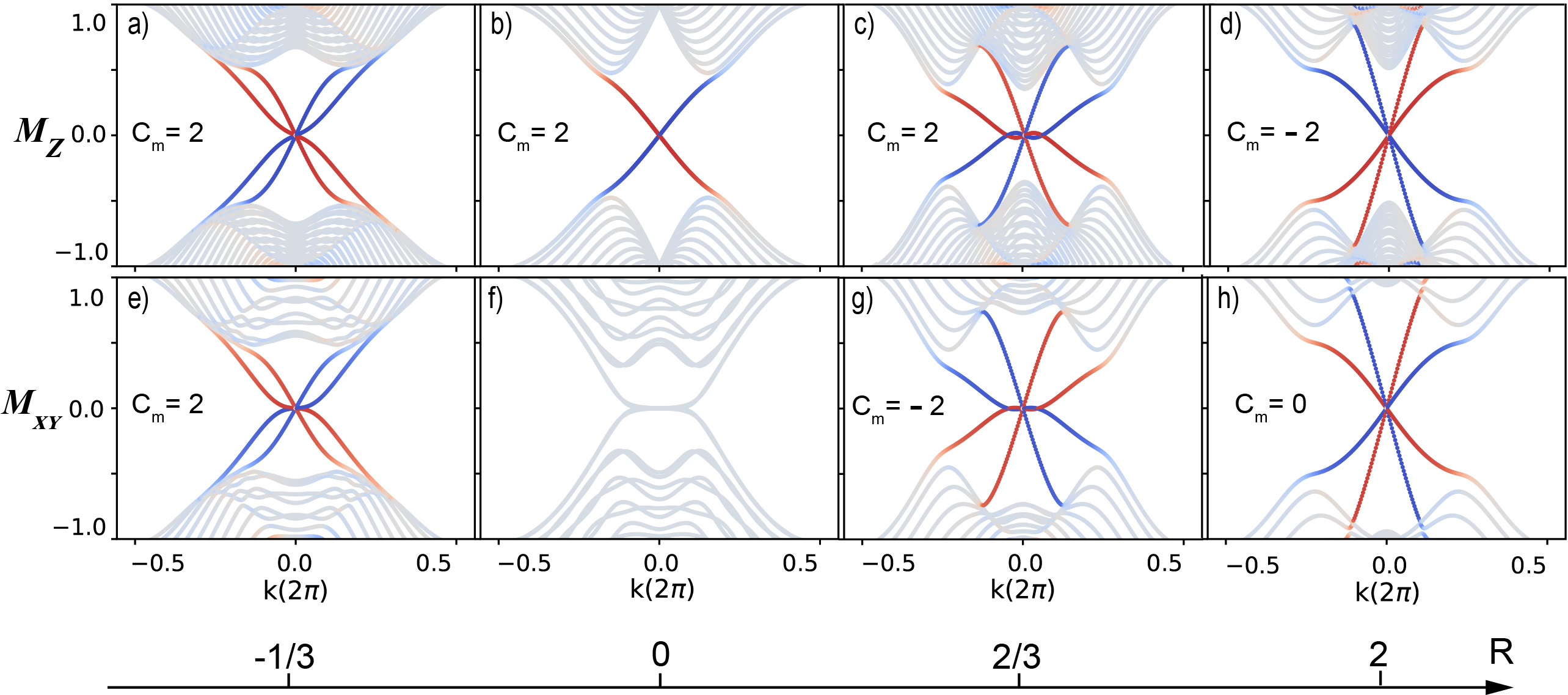}
\caption{Surface states for the $+i$ subspace in $M_{z}$ and $M_{xy}$ invariant planes for different values of $R$. These plots are made from the two-dimensional slab models for the $+i$ $M_{z}$ or $M_{xy}$ subspace of the bulk Hamiltonian (\ref{Hameqn});
the four figures in the top row are surface states in the $+i$ subspace of $M_{z}$, while the four figures in the bottom row are surface states in the $+i$ subspace of $M_{xy}$. 
The horizontal axis is along the corresponding symmetry-preserving line in the surface Brillouin zone.
In a slab configuration, the blue bands are localized at the front edge and the red points are localized at the other edge. Without loss of generality, we can focus on the states with blue color.
In (a) and (e), $R=-1/3$; in (b) and (f), $R=0$, in (c) and (g), $R=2/3$; in (d) and (h), $R=2$. Figure (f) shows a phase transition for only the $M_{xy}$ surface states as discussed in Appendix~\ref{mxy}. Figure (h) shows a trivial state with $C_m=N^+_R-N^+_L=0$. The crossing at $\Gamma$ can be gapped by adding the quadratic terms that reduce the symmetry from ${SO}(3)$ to $Pm\bar{3}m$, as discussed at the end of Appendix~\ref{mxy} and shown in Fig.~\ref{figbeta}.}
\label{figHplus}
\end{figure*}

If $0<R<1$, we can get the energy eigenvalues in the same way. 
The slopes for the $+i$ subspace are $(\frac{v_s}{2}\pm v_d)k_x$. If $0<R<1/2$, the slopes are both positive, corresponding to mirror Chern number $C_m=2$. If $1/2<R<1$, the two slopes have opposite signs, but the mirror Chern number cannot change because $R=1/2$ does not correspond to the bulk band gap closing. Instead, the surface state is a cubic-like curve that crosses the Fermi level three times, which we have verified numerically. Fig.~\ref{figHplus}(c) shows one example when $\frac12<R<1$. Focusing on the blue curves, indicating localization on the top boundary, the mirror Chern number is equal to number of right-moving bands ($N_R$) minus left-moving bands ($N_L$): if we count the bands at energy 0.5, $N_R=2$, $N_L=0$, while if we count at energy 0, $N_R=3$, $N_L=1$. Thus, there is no contradiction and we deduce $C_m=N_R-N_L=2$.\par
Fig.~\ref{figHplus} shows examples of the surface state spectrum for the $+i$ sector as $R$ is tuned through phase transitions (according to (\ref{eqncn}), analyzing $+i$ subspace is enough to determine the mirror Chern number.)

\par

\begin{figure}
    \centering
    \includegraphics[width = 5cm]{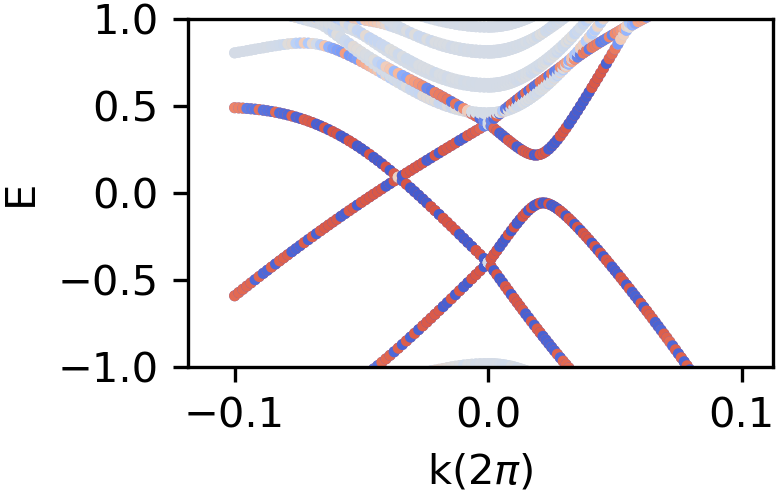}
    \caption{Adding quadratic terms is necessary to gap the surface states in regions where the mirror Chern number is trivial. Here we use the parameters $m=-1$, $\alpha_1=-\alpha_2=5$, $\beta_1=\beta_2=2$, $\gamma_1=\gamma_2=0$, $v_d=2$, $v_s=1$. $k$ is plotted along a segment of the path $\bar{X}-\bar{\Gamma}-\bar{M}$. In this case $R = 2$, $C_{m}(M_z)=-2$, and $C_{m}(M_{xy})=0$. The trivial surface states along $\bar{\Gamma}-\bar{M}$ are gapped by adding the $\beta$ terms.}
    \label{figbeta}
\end{figure}

\subsubsection{Surface potential}

We now derive the surface theory in the presence of a surface potential.
As discussed in the main text, the surface potential can result from different sources, such as cleaving to reveal a low-symmetry surface, making a heterostructure, or adding strain. Regardless of the source, we model these effects by adding the lowest order $M_z$-breaking surface potential:
\begin{align}
    V_{surface} = & \begin{bmatrix}
    V(\delta_{a1},\delta_{b1}) &
   0 \\
   0&
  V(\delta_{a2},\delta_{b2})
   \end{bmatrix}
   \\
   V(\delta_a,\delta_b) = & ~~ \delta_a \sigma_x\tau_z - \delta_b \sigma_y\tau_z,
\end{align}
which is the same surface potential as Eq.~(\ref{eqn_surface_potential}), taking $\delta_c=\delta_d=0$.
\par

We first re-express this surface potential in basis of $M_z$ eigenstates, then project it to the low-energy subspace spanned by $\langle \psi_1,\psi_2,\psi_3,\psi_4 \rangle$, where $\psi_1,\psi_2$ are the zero-energy wavefunctions in the $+i$ subspace, and $\psi_3,\psi_4$ are the zero-energy wavefunctions in the $-i$ subspace. For the $v_d>v_s>0$ case, the linear order Hamiltonian becomes:
\begin{align}
    H_{eff} = & v_d k_x \sigma_0\tau_z -\frac{v_s}{2} k_x \sigma_z\tau_z +\frac{\sqrt{3}}{2}v_s k_x \sigma_x\tau_0 \nonumber\\
    & ~~ + \frac12 (\delta_{a1}+\delta_{a2})\sigma_x\tau_x+\frac12(\delta_{b1}-\delta_{b2})\sigma_x\tau_y
\end{align}
The four energy eigenvalues for this effective surface model are:
\begin{equation}
    E = \pm v_s k_x \pm \sqrt{v_d^2 k_x^2+(\frac{\delta_{a1}+\delta_{a2}}{2})^2+(\frac{\delta_{b1}-\delta_{b2}}{2})^2 }
\end{equation}
Since the term under the square root is positive definite, a surface gap is always opened with this term.
It is worth mentioning that the surface potential preserves inversion symmetry and time reversal symmetry, so the hinge states should not be gapped. Similar results for $M_x$ and $M_y$ surface states can be derived in the same way. Fig.~\ref{figsp} shows the $M_x$ and $M_y$ surface states become gapped upon adding certain surface potentials.

\subsection{$M_{xy}$ mirror Chern surface states}
\label{mxy}

It is useful to first define $k_1 = (k_x+k_y)/\sqrt{2}$, $k_2 = (k_x-k_y)/\sqrt{2}$. The mirror invariant plane then is $k_2 =0$.
The transformation matrix that block diagonalizes the Hamiltonian in this plane, found from the eigenstates of $C_{2,1\bar{1}0}{\cal  P}$, is given by
\begin{equation}
    U =
    \left(
    \begin{smallmatrix}
 0 & 0 & \frac{1}{2}-\frac{i}{2} & 0 & 0 & 0 & -\frac{1}{2}+\frac{i}{2} & 0 \\
 0 & 0 & 0 & -\frac{1}{2}-\frac{i}{2} & 0 & 0 & 0 & \frac{1}{2}+\frac{i}{2} \\
 0 & 0 & 0 & \frac{1}{\sqrt{2}} & 0 & 0 & 0 & \frac{1}{\sqrt{2}} \\
 0 & 0 & \frac{1}{\sqrt{2}} & 0 & 0 & 0 & \frac{1}{\sqrt{2}} & 0 \\
 -\frac{1}{2}+\frac{i}{2} & 0 & 0 & 0 & \frac{1}{2}-\frac{i}{2} & 0 & 0 & 0 \\
 0 & \frac{1}{2}+\frac{i}{2} & 0 & 0 & 0 & -\frac{1}{2}-\frac{i}{2} & 0 & 0 \\
 0 & \frac{1}{\sqrt{2}} & 0 & 0 & 0 & \frac{1}{\sqrt{2}} & 0 & 0 \\
 \frac{1}{\sqrt{2}} & 0 & 0 & 0 & \frac{1}{\sqrt{2}} & 0 & 0 & 0 \\
    \end{smallmatrix}
    \right)
\end{equation}

In this basis, the linear order Hamiltonian again decomposes into $H_{M_{xy}}^{+i}\oplus H_{ M_{xy}}^{-i}$, where
\begin{align}
\label{ham_mxy}
    H_{M_{xy}}^{\pm i} =& -m\tau_z +\left( \sqrt{\frac{3}{8}}v_s(\sigma_x-\sigma_y)\tau_x \mp(v_d-\frac{v_s}{2}\sigma_z)\tau_y\right)k_1 \nonumber \\
    &-(v_d\sigma_z+v_s)\tau_x k_z
\end{align}

Following the same steps as we did to derive the surface states protected by $M_z$, we consider a sample that is finite in the $k_1$ ($w_1$) direction and replace $k_1$ with $-i\partial_{w_1}$, where $w_1=(x+y)/\sqrt{2}$, $w_2=(x-y)/\sqrt{2}$. Now $k_2$ and $k_z$ are good quantum numbers and $k_2=0$ is the mirror-invariant plane we are interested in. We take the mass to be a function of $w_1$, i.e.,  $m(w_1)$, and model the boundary by allowing it to change sign when $w_1$ changes sign. We start with ansatz: $\Psi = {\cal  N}e^{-\frac{1}{\xi}\int_0^{w_1}m(w_1)dw_1} \chi$. Solving the Schr\"{o}dinger equation when $k_z=0$ in Eq.~(\ref{ham_mxy}) yields two physical solutions for $+i$ sector. The first one is $\xi_1 = \frac12 v_s+ \sqrt{v_d^2+\frac34v_s^2}$, with spinor
\begin{equation*}
\scriptstyle
  \chi_1 =
\scriptstyle 
  N_1\left( \sqrt{\frac 23}(1-i)\frac{\sqrt{v_d^2+\frac34 v_s^2}-v_d}{v_s}, ~ 1 ~, -\sqrt{\frac 23}(1-i)\frac{\sqrt{v_d^2+\frac34 v_s^2}-v_d}{v_s}, ~ 1\right),
\end{equation*}
where $N_1 = 1/\sqrt{2+\frac83((\sqrt{v_s^2+\frac34 v_d^2}-v_d)/v_s)^2}$. 
The other solution is $\xi_2 = -\frac12 v_s+ \sqrt{v_d^2+\frac34v_s^2}$, and the spinor is
\begin{equation*}
\scriptstyle
  \chi_2 =
\scriptstyle 
  N_2\left( \sqrt{\frac 23}(1-i)\frac{\sqrt{v_d^2+\frac34 v_s^2}+v_d}{v_s}, ~ -1 ~, \sqrt{\frac 23}(1-i)\frac{\sqrt{v_d^2+\frac34 v_s^2}+v_d}{v_s}, ~ 1\right),
\end{equation*}
where $N_2 = 1/\sqrt{2+\frac83((\sqrt{v_s^2+\frac34 v_d^2}+v_d)/v_s)^2}$. 
\par

We now project $H_{ M_{xy}}^{+i}$ onto the low-energy subspace spanned by $\chi_1,\chi_2 $, and then solve for the energy eigenvalues. We find that in the $+i$ subspace, the $M_{xy}$ mirror Chern surface states have the following linear band dispersion:

\begin{equation}
    E_1 = k_z v_d\left(1-\frac{v_s}{\sqrt{v_d^2+\frac34v_s^2}}\right)
\end{equation}
\begin{equation}
    E_2 = -k_z v_d\left(1+\frac{v_s}{\sqrt{v_d^2+\frac34v_s^2}}\right)
\end{equation}

The two signs of the slopes vanish at $R=v_d/v_s=0$, and one of them vanishes at $R=\pm 1/2$. Again there is no phase transition at $R=\pm1/2$ because the bulk gap is not closed. 
Numerical calculation for $+i$ subspace (Fig.~\ref{figHplus}(c)) shows that the surface states on the top boundary (color red) have a cubic dispersion, and the mirror Chern number is counted to be 2 in the same way as we did for $M_z$. We conclude: when $0<R<1/2$, both slopes are negative and the mirror Chern number is $-2$; when $1/2<R<1$, the two slopes have opposite sign but nonetheless the mirror Chern number is unchanged ($C_m=-2$) 
;  when $-1/2<R<0$, both slopes are positive and the mirror Chern number is $C_m=2$; when $-1<R<-1/2$, again the two slopes have opposite sign, but the mirror Chern number is still $C_m=2$.\par

For the case $R>1$ or $R<-1$, numerical calculation (some $R$ values are shown in Fig.~\ref{figHplus}) shows that they differ from the case $1/2<R<1$ in that the surface bands no longer have the cubic function shape. The surface bands now don't change slope at finite $k$, although the zero-energy states near the origin share the same formula. When $R=1$, the bulk gap closes at some non-zero points along $\Gamma-X$, which
changes the mirror Chern number by 2. \par

We now comment on the linear $k.p$ model presented in Ref.~\onlinecite{hsieh2014topological}.
In the linear model, gapless surface states appear even in trivial phases, as well as along low-symmetry $k$ directions that do not have a mirror symmetry.
This is because there is always a four-fold surface band crossing at $\Gamma$ in phase (I) as a result of the ${SO}(3)$ rotation symmetry in the linear model (as well as the fact that there is always some direction with a nontrivial mirror Chern number in phase (I); see Table~\ref{table3}). 
However, if we  add the quadratic terms in the Hamiltonian (\ref{Hameqn}), the symmetries will be reduced to $Pm\bar{3}m$, allowing the surface band gap to open. This illustrates why a linear $k.p$ model is insufficient to describe the surface states of the antiperovskites. \par

We perturbatively consider the quadratic term $H_{quadratic}$ (\ref{eqn_quadratic}). Let us consider the $+i$ subspace with the condition $v_s=0$. 
The surface states have energies $\pm v_d k_z$ and the mirror Chern number is zero. The eigenstates for $+i$ sector are $\psi_j={\cal N}e^{-\frac{1}{v_d}\int_0^{w_1}m(w_1)dw_1}\chi_j$, where $\chi_1=\frac{1}{\sqrt{2}}(1,0,1,0)$, $\chi_2=\frac{1}{\sqrt{2}}(0,1,0,1)$. 
We denote the eigenstates as $\psi_{1,2}$. The first order energy correction from the quadratic term is: $\langle \psi_{j}|H_{quadratic}|\psi_{j}\rangle$. The energy difference of the two eigenstates in the $+i$ subspace at $\Gamma$ is 
\begin{equation}
\label{eqn_dE}
    \Delta E = -\frac{1}{4}(\beta_1+\beta_2)\langle \psi_{j}|\partial^2_y|\psi_{j}\rangle.
\end{equation}
If we choose $m(w_1)=m_0\tanh(w_1)$, then there is an analytical result for this expression in the case $v_s=0$:
\begin{equation}
    \Delta E = \frac14(\beta_1+\beta_2)\frac{3(m_0/v_d)^2}{2m_0/v_d+1}.
\end{equation}

This energy difference shifts the two $+i$ bands, lifting the four-fold degeneracy at $\Gamma$ to two two-fold degeneracies. Now the crossing point of the two $+i$ bands has been moved away from the origin. Let's denote the point along $\Gamma-X$ as $k_0(M_z)$ and the point along $\Gamma-M$ as $k_0(M_{xy})$. 
It is now possible to open a gap at $k_0(M_{xy})$, while $k_0(M_z)$ can remain gapless; this was not possible in the $SO(3)$ symmetric model. Fig.~\ref{figbeta} shows that adding $\beta \neq 0$ can open a surface gap for $M_{xy}$ surface states when $R=2$ (where the $M_{xy}$ mirror Chern number is zero).\par


\section{\label{appc}$M_{xy}$ and $M_{\bar{x}y}$ hinge states }
{
In addition to the HOTI phase protected by inversion symmetry, indicated by non-trivial $\mathbb{Z}_4$ index, antiperovskites are also in the HOTI phase protected by mirror symmetries and time reversal symmetry, indicated by even mirror Chern numbers (similar to SnTe~\cite{hsieh2012topological,schindler2018higher}).
}
\par
{
For the same reason that surface mirror Chern states obscure the hinge states, most of the mirror symmetries should be broken to open the surface gaps. For the rod geometry with $(100)$ and $(010)$ surfaces, gapless hinge states appear when $M_{xy}$ and $M_{\bar{x}y}$ are preserved while other mirror symmetries are broken. Notice that preserving $M_{xy}$, $M_{\bar{x}y}$ and inversion symmetry at the same time implies preserving $M_z$ symmetry. }
{
Thus, in order to break $M_z$ symmetry -- necessary for gapping the surface states -- it is necessary to break either $M_{xy}$, $M_{\bar{x}y}$ or inversion. If inversion symmetry is broken and both $M_{xy}$ and $M_{\bar{x}y}$ are preserved, then there will be hinge states on all four hinges.
Here, as an example, we preserve inversion symmetry and break either $M_{xy}$ or $M_{\bar{x}y}$, resulting in a configuration with hinge states on two of the four hinges.}
\par
\begin{figure}[h]
    \centering
    \includegraphics[width=8cm]{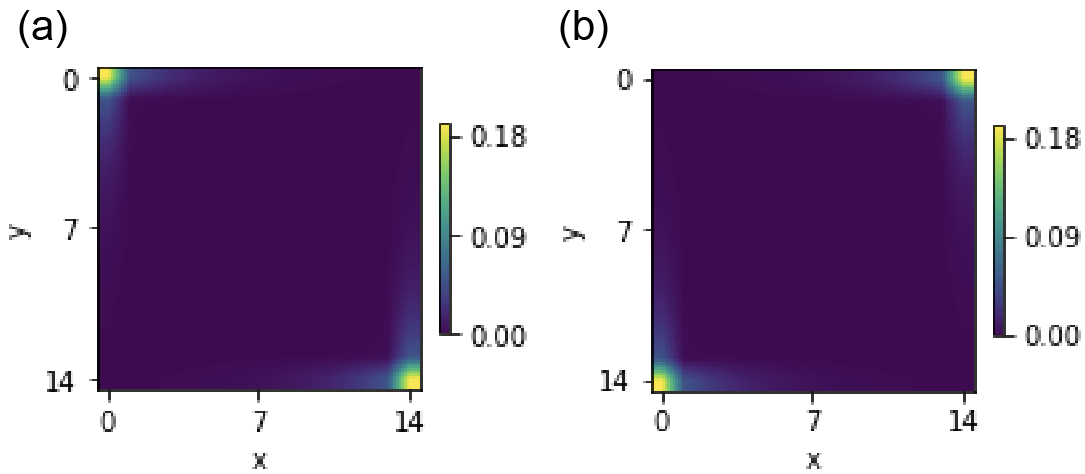}
    \caption{The same strain terms $V_1=\delta_{a1}=-\delta_{a2}$ and $V_2=\delta_{b1}=-\delta_{b2}$ are used. (a) As shown in Table~\ref{table4}, $V_1=-V_2=0.1$ preserves only $M_{\bar{x}y}$, while the other eight mirror symmetries are broken. Hinge states are now localized at left-top and right-bottom corners. (b) When $V_1=V_2=0.1$, only $M_{xy}$ is preserved. Hinge states are now localized at left-bottom and right-top corners.}
    \label{fig_appc}
\end{figure}

{
The real space distributions of these gapless states at $k_z=0$ are shown in Fig.~\ref{fig_appc}. Although these mirror protected hinge states can be obtained in theory, it may be harder to realize them in experiment because it requires that the two strain parameters satisfy $\delta_a=\pm\delta_b$, {i.e., it requires fine-tuning}. When the parameters do not obey this relation, these ``mirror Chern hinge states'' should be interpreted as the inversion protected hinge states. When $\delta_a=0$ or $\delta_b=0$, the surface gap closes on the $(010)$ or $(100)$ surface respectively, allowing for a phase transition between the two configurations of hinge states.
}

\end{document}